\documentstyle[12pt]{article}
\advance\textheight by \topskip
\begin{document}
\newcommand{\Psl}{\not\!\! P}
\newcommand{\dsl}{\not\! \partial}
\newcommand{\half}{{\textstyle\frac{1}{2}}}
\newcommand{\for}{{\textstyle\frac{1}{4}}}
\newcommand{\eqn}[1]{(\ref{#1})}
\def\la{\mathrel{\mathpalette\fun <}}
\def\a{\alpha}
\def\b{\beta}
\def\g{\gamma}\def\G{\Gamma}
\def\d{\delta}\def\D{\Delta}
\def\e{\epsilon}
\def\et{\eta}
\def\z{\zeta}
\def\t{\theta}\def\T{\Theta}
\def\l{\lambda}\def\L{\Lambda}
\def\m{\mu}
\def\f{\phi}\def\F{\Phi}
\def\n{\nu}
\def\p{\psi}\def\P{\Psi}
\def\r{\rho}
\def\s{\sigma}\def\S{\Sigma}
\def\ta{\tau}
\def\x{\chi}
\def\o{\omega}\def\O{\Omega}
\def\lagr{{\cal L}}
\def\cd{{\cal D}}
\def\k{\kappa}
\def\tz{\tilde z}
\def\tF{\tilde F}
\def\ri {\rightarrow}
\def\cf{{\cal F}}
\def\pa {\partial}
\begin{flushright}
SU-ITP-92-13 \\ \today
\end{flushright}
\vspace{-1.2cm}
\begin{center}
{\large\bf SUPERSYMMETRY AS  A COSMIC CENSOR}\\
\vskip .9 cm
{\bf Renata Kallosh \footnote {On leave  from: Lebedev
Physical Institute, Moscow. \  Bitnet address:
kallosh@slacvm},
Andrei Linde \footnote{On leave  from: Lebedev
Physical Institute, Moscow.   \  E-mail:
linde@physics.stanford.edu},
Tom\'as Ort\'in \footnote{Bitnet
address: tomaso@slacvm}
 \vskip 0.01cm
 Amanda Peet  \footnote{Bitnet address: peet@slacvm} and
 Antoine Van Proeyen\footnote
{Permanent address: Instituut voor Theoretische Fysica,
Universiteit Leuven,
B-3001 Leuven, Belgium.\\ \indent \ \ \ Onderzoeksleider, N.F.W.O.
Bitnet address: fgbda19@blekul11}}
 \vskip 0.05cm
Physics Department, Stanford University, Stanford   CA 94305
\end{center}
\vskip .6 cm
\centerline{\bf ABSTRACT}
\vspace{-0.7cm}
\begin{quote}
\ \ \ \ \ In supersymmetric theories the mass of any state is bounded
below by the values of some of its charges. The corresponding bounds
in case of Schwarzschild ($M \geq 0$ ) and Reissner-Nordstr\"om
($M \geq |q|$) black holes are known to coincide with the
requirement that naked singularities be absent.

 \ \ \ \ \  Here we investigate $(U(1))^2$ charged dilaton black
holes in this
context.   The extreme solutions are shown to saturate the
supersymmetry
bound of $N=4\ d=4$ supergravity, or dimensionally reduced
superstring
theory. Specifically, we have shown that extreme dilaton black holes,
with
electric and magnetic charges, admit super-covariantly constant
spinors.
The supersymmetric positivity bound for dilaton black holes,
$M \geq \frac{1}{\sqrt 2} \,( |Q|+|P|)$, takes care  of the absence
of naked
singularities  of the dilaton black holes and is, in this sense,
equivalent to
the cosmic censorship condition.

\ \ \ \ \ The temperature, entropy and singularity of the stringy
black hole
are discussed in connection with the extreme limit and  restoration
of
supersymmetry. The Euclidean action (entropy) of the extreme black
hole
is given by $2\pi |PQ|$.  We argue that this result is not altered by
higher
order corrections in the supersymmetric theory.  In Lorentzian
signature,
quantum corrections to the effective on-shell action in the extreme
black hole
background are also absent.

\ \ \ \ \  When a black hole reaches its extreme limit, the thermal
description breaks down. It cannot continue to evaporate  by emitting
(uncharged) elementary particles, since this would violate the
supersymmetric positivity bound.  We speculate on the possibility
that  an extreme black hole may ``evaporate" by emitting smaller
extreme black holes.

\end{quote}

\normalsize
\newpage

\section{Introduction }

Evaporation of black holes is one of the most interesting effects of
nonperturbative quantum gravity. Investigation of this process may
help us to
understand the nature of singularities in gravitational theory, the
problem of
information loss during the process of black hole evaporation and the
interplay between quantum and thermal descriptions of processes near
black holes. All of these problems become especially urgent in the
theory
of the last stages of black hole evaporation. For a Schwarzschild
black hole,
this
happens when its mass approaches the Planck value $M_p$. In this case
any
semiclassical description of the black hole becomes impossible.
Therefore,
despite many attempts, we still do not have a complete understanding
of the
last stages of black hole evaporation.

Recently there have been many new attempts to study quantum effects
near
black holes. Most of them are related to more complicated black
holes, such as
magnetically or electrically charged black holes, dilaton black
holes, etc.
First
of all, this provides a more complete picture of black hole physics
in the
context of theories of elementary particles and/or superstrings.
Moreover,
some aspects of the theory of such black holes prove to be simpler
than the
corresponding aspects of ordinary black holes. For example,
evaporation
of a charged Reissner-Nordstr\"om     black hole stops when its mass,
measured in units of the Planck mass $M_p$,
 approaches the absolute value of its  charge  $|q|$. Thus, for a
sufficiently large charge, one may study the last stages of black
hole
evaporation  when the mass of the black hole is much larger than
$M_p$
and quantum fluctuations of the metric are not too
strong.\footnote{Throughout \, this paper we will work in a system of
units
where $M_p = 1$.}

However,  strongly charged black holes may discharge by creation of
pairs of charged elementary particles. To isolate these effects from
the
effects of quantum gravity, in which we are mainly interested, one
may
consider models without charged elementary particles. In such models
electric and magnetic fields are not produced by charged particles,
but
flow from infinity or from the singularity.  At the classical level,
there
is no difference between a Reissner-Nordstr\"om black hole with
charged particles  in its center and a black hole with a spherically
symmetric
electric field originating from the singularity. In both cases we
will have
the same theory of black hole evaporation, but in the last case we do
not
need to be concerned with extra complications, such as  particle
production
in strong  electric fields, charge quantization, etc.

This idea proves to be especially productive if one can find a way of
embedding the original bosonic theory in a  supersymmetric  theory.
In
such a case one has the same description of the classical properties
of black
hole and of  its evaporation, but higher order quantum corrections
are under
much better control. For example,  an extreme  Reissner-Nordstr\"om
black
hole with mass $M = |q|$ may be embedded in  $N=2$ supergravity
\cite{GH},
\cite{GG}.   All higher order quantum  corrections to the effective
action of
$N=2$ supergravity  in the field of the extreme black hole  could be
shown
to vanish if the theory had no anomalies \cite{K}. In particular, the
effective
action would have no imaginary part, which would mean  no
particle creation in the field of an extreme Reissner-Nordstr\"om
black hole. These and some other properties of
supersymmetric black holes indicate that they may prove to be  a
unique
laboratory for investigation of black hole physics and quantum
gravity in
general.  Indeed, until we started studying supersymmetric black
holes, we
had no example  of  a nontrivial  Lorentzian  four-dimensional
background
where all quantum gravity corrections to the effective action
vanished.

The theory of $N=2$ supergravity,  however,  has one-loop anomalies
\cite{Duff}. Therefore the formal proof of the absence of quantum
corrections for the classical extreme  Reissner-Nordstr\"om  black
hole
is not sufficient.
One is led to  try to find a supersymmetric embedding  of charged
dilaton
black holes  \cite{G} - \cite{HW}   in   dimensionally reduced string
theory or
$d=4, N=4$ supergravity, where the anomalies can  be cancelled.

Surprisingly enough, until now, no such supersymmetric embedding has
been found.   It was argued in \cite{GG} that the lower bound on the
mass of
such black holes  does not follow from  supersymmetry, as the
supersymmetric theory investigated in \cite{GG} was a Kaluza-Klein
compactification of five-dimensional supergravity. In \cite{GHS} it
was
found that there was again no suitable embedding if one takes the
vector
field to be in a Yang-Mills multiplet.   Moreover,  supersymmetry is
usually
related to zero temperature.
However,  the thermal properties of dilaton black holes are somewhat
unusual,
(the corresponding literature contains several contradictory
statements on this
issue, see e.g. \cite{GM} - \cite{HW}), and there was no clear signal
for
supersymmetry from the temperature, unlike the extreme
Reissner-Nordstr\"om case where $T = 0$.

Such dilaton black holes, if supersymmetrically embedded in
supergravity
and/or superstring theory, would  be  especially interesting since
they might
lead to new insights into perturbative and non-perturbative quantum
effects
in these theories.

In this paper we will investigate dilaton black holes in this
context.  We will
consider $U(1)~\otimes~U(1)$ dilaton black holes,  with electric
and magnetic charges and without
axion. They do not coincide with dual dilaton dyons \cite{STW}, which
are dual
rotations of the purely electric or purely magnetic charged dilaton
black
holes.
Under that dual rotation, the second charge arises together with the
axion, but
the  metric, causal structure and thermodynamic properties of the
dual dilaton
dyon are the same as in the purely electric or purely magnetic
solution without
axion. In our solution, the existence of a second charge does change
the
metric, causal structure and thermodynamic properties of the
solution. The
importance of the second charge is related to the existence of two
central
charges in $N=4$ supersymmetry. In
particular, we will show that {\it extreme dilaton black holes  are
supersymmetric in the context of $N=4$ supergravity}. We will also
show that
the lower bound on the dilaton black hole mass imposed by cosmic
censorship,

\begin{equation}\label{b}
M^2 +\S^2 \geq P^2 + Q^2 \ , \\
\end{equation}

\noindent does coincide with the bound which can be derived from
supersymmetry.
In equation (\ref{b})
$\S, P, Q $ are the dilaton, magnetic and electric charges,
respectively.
Equality in (\ref{b}) shows that, at the extremal value of the mass,
supersymmetry leads to the balance of   gravitational,
electromagnetic
and dilatonic forces.  Supersymmetric dilatonic {\it multi} black
hole
solutions,
satisfying the force-balance  condition, will be exhibited.

The positivity bound (\ref{b}) implies, in particular, that the black
hole
singularity remains hidden by  the event horizon until  the mass of
the black
hole decreases to its extreme value. If the black hole has both
electric
and magnetic charge, the singularity always remains hidden by the
event
horizon. It
approaches the horizon only if the extreme dilaton black hole has
purely
electric
or purely magnetic charge. Even in this case, though, any external
observer who
does not touch the singularity cannot see it.  In this sense,
supersymmetry
 plays the role of a cosmic censor. It keeps the singularity away
from the
eyes of any observer who does not want to fall into the black
hole.

 The paper is organized as follows.
In Sec. 2 we describe the relation between the supersymmetric
positivity
bound and cosmic censorship for classical Schwarzschild and
Reissner-Nordstr\"om black holes.
In Sec. 3 the spherically symmetric electrically and magnetically
charged  dilaton black
 hole is presented as a solution of dimensionally reduced superstring
theory.  This solution includes (for some particular values of
electric, magnetic and dilaton charges) the classical Schwarzschild
and
Reissner-Nordstr\"om  black holes and dilaton black holes with
either purely electric, purely magnetic, or both charges present in
the
solution.

In Sec. 4 the extreme multi black holes are described, as well as
spherically symmetric  electrically and magnetically charged
extreme dilaton black
holes. It is also explained that the  purely electric  extreme
dilaton black
holes
are special cases of a metric of
Bonnor  \cite{B} describing charged dust in equilibrium
\footnote{The purely
magnetic ones also fit in to this category, as can be seen by
performing a
duality transformation.}. These metrics generalize the
Papapetrou-Majumdar class of metrics in the presence of specific
sources in
the Einstein and Maxwell equations called "charged dust".
The
Papapetrou-Majumdar metrics are supersymmetric extreme multi black
hole
solutions \cite{GH} of Einstein-Maxwell theory.

In Sec. 5 the thermal properties of the dilaton black holes are
discussed, in particular, the temperature, entropy and specific heat.
The extreme dilaton black holes with $PQ\neq 0$ are shown to have
zero temperature, whereas their entropy is given by $2\pi |PQ|$. It
is
shown that the thermal description
of stringy black holes near extremality breaks down for all possible
values of
the charges $P$ and $Q$.

Sec. 6 contains an investigation of the supersymmetric properties
of dilaton black holes, in the context of $N=4$ supersymmetry.
It is shown that non-extreme dilaton black holes necessarily break
all
supersymmetries. The unbroken $N=1$
supersymmetries of electrically and magnetically charged extreme
multi black holes are found. In addition,
the unbroken $N=2$ supersymmetries of purely electric and purely
magnetic
extreme multi black holes are identified.

In Sec. 7 the partition function of the dilaton black holes is
calculated in
the classical approximation. A nonrenormalization theorem for quantum
corrections to the extreme dilaton black hole partition function is
outlined.

In Appendix A we introduce our conventions and compare them with
those used by other authors. In Appendix B  some speculative ideas
about
splitting of extreme dilaton black holes are presented.

In the Figures, we plot different characteristics of the charged
dilaton black
hole, such as temperature and entropy, by using the program
"Mathematica".
\vskip 1 cm

\section{Supersymmetric Positivity Bound and Cosmic Censorship}
In order to make our goals and methods more clear, let us remember
some
basic facts about ordinary Schwarzschild black holes with metric
\begin{equation}\label{1}
ds^2 = \left(1 - \frac{2M}{r}\right)\, dt^2 -  \left(1 -
\frac{2M}{r}\right)^{-1} dr^2 - r^2 d\O \ .
\end{equation}
For $M > 0$, this metric has a singularity at $r_g = 2M$. However,
this is just
a coordinate singularity, which corresponds to the event horizon
where the
components of the metric $g_{tt} = g^{-1}_{rr} = (1 - \frac{2M}{r})$
change their
sign. The true singularity, where the curvature tensor becomes
infinite,
is at $r = 0$.

The presence of singularities, i.e. of places where the normal laws
of
physics formulated in terms of classical space-time break down,
is one of the main problems of classical general relativity. However,
in many cases this problem is somewhat softened. For example, any
observers near the Schwarzschild black hole  cannot actually {\it
see}
any violation of
standard laws of physics until they reach the singularity. Indeed,
the change
of sign of $g_{tt} = g^{-1}_{rr}$ inside the black hole means that
the
light cone
inside it
looks inwards (T - region) \cite{PenHawk}. It is possible to send a
signal
towards the singularity, but it is impossible to get any signal
backwards.
Therefore, we will  not have information about the singularity until
we fall into it, and then we will not care.

The situation with electrically charged Reissner-Nordstr\"om black
holes with
metric
\begin{equation}\label{2}
ds^2 = \left(1 - \frac{2M}{r} + \frac{q^2 }{r^2}\right)\, dt^2 -
\left(1 -
\frac{2M}{r} +
\frac{q^2 }{r^2}\right)^{-1} dr^2 - r^2 d\O \
\end{equation}
is slightly more complicated. When the mass $M$ of the black
hole is
larger than the absolute value of its  charge, $|q|$, the singularity
at $r =
0$ is hidden from us by two  horizons located at
\begin{equation}
r_+ = M + \sqrt{M^2 - q^2 } \ , \ \
r_- = M - \sqrt{M^2 - q^2 } \ .
\end{equation}
At each of these horizons $g_{tt}$ changes its sign. An observer
deciding to fly to the region $r < r_-$ would be able to see the
singularity.
However, an outside observer staying  at $r > r_-$ cannot see the
singularity
for the same reason
as in the Schwarzschild space: the region between $r_-$ and
$r_+$ is a T - region, where all signals can go only towards smaller
$r$.

The extreme case $M = q$ is special. In this case the two horizons
$r_+$ and
$r_-$ coincide and the T - region between them disappears. But even
in this
case an outside observer staying at any finite distance from the
horizon $r_+ = r_- = M$ will not see the singularity. Moreover, just
as in  the
Schwarzschild space, the observer will not see anything which is
hidden under
the horizon {\it or coincides with the horizon}. Indeed,
the equation describing the radial motion of a wavefront of
light is $ds = 0$, i.e.
\begin{equation}
\left({dr\over dt}\right)^2 = {g_{tt}\over g_{rr}}\ .
\end{equation}
The time taken for a signal to go from $r_1$ to $r_2$ is
given by \begin{equation}\label{3}
t = \int_{r_1}^{r_2} \sqrt{g_{rr}\over g_{tt}}\, dr \ .
\end{equation}
One can easily check that this time diverges if $r_1$ coincides with
$r_g =
2M$ for the Schwarzschild black hole or with $r_+$ for the
Reissner-Nordstr\"om black hole. Thus the horizon at $r_g$ (at $r_+$)
is called an {\it event horizon}: Our part of the universe cannot be
influenced
by any event which may happen in a region covered by the event
horizon
or coinciding with it. In particular, we cannot see a singularity
if it is covered by (or coincides with) the event horizon.

Thus, in the Schwarzschild space the singularity cannot be seen from
any place
with $r > 0$. In the Reissner-Nordstr\"om space the singularity can
be seen
from a place with  $r < r_-$, but it cannot be seen from any place
with $r \geq
r_-$. However, in a Schwarzschild metric with $M < 0$, as well as in
the
Reissner-Nordstr\"om metric with $M < |q|$, the singularity does not
coincide
with any horizon, is not hidden by any horizon and, therefore, is
visible from
any place. There exists a cosmic censorship conjecture, which says
that naked
singularity of such type cannot be formed. There are several versions
of this
conjecture, which differ from each other by specific assumptions
concerning
initial conditions for gravitational collapse and the structure of
the energy
momentum tensor  (weak cosmic censorship conjecture \cite{Pen1},
strong
cosmic censorship conjecture \cite{Pen2}; see, e.g., \cite{Wald}).
Even the
definition of a naked singularity is author-dependent: Is the
singularity
naked if it can be seen from the horizon? What if the horizon itself
is
singular? To avoid unnecessary complications, we will say that the
singularity is hidden if it is covered by (or coincides with) the
event
horizon,
i.e. if  an observer staying at any finite distance from the horizon
cannot see
the singularity.

A general proof of the cosmic censorship conjecture is still absent,
and
several (rather artificial) exceptions are known. It is very
interesting,
therefore, that for some supersymmetric theories to be discussed
below, the
cosmic censorship conjecture in the form mentioned above coincides
with
the supersymmetric positivity bound, which requires that the mass of
the
asymptotically flat space-time be larger than or equal to the
absolute values
of all central charges.

In extended global supersymmetry, the mass of any quantum state is
bounded
 below by the moduli of the eigenvalues of the central
charges $z_n$ of the supersymmetry algebra with $N$ spinor
operators, $n= 1,2,\dots \frac{N}{2}$  \cite{WO,FSZ}. Consider,
specifically,
$N=4$ theory with two central charges $z_1, z_2$. From the SUSY
algebra
in the rest frame it can be derived that

 \begin{eqnarray}
 \{S_{(1)} , S_{(1)}^{*}\}&=& 2|S_{(1)}|^2 = M - |z_1|  \geq
0 \ , \nonumber\\
 \{S_{(2)} , S_{(2)}^{*}\}&=& 2|S_{(2)}|^2 = M - |z_2|  \geq
0 \ ,\nonumber\\
  \{T_{(1)} , T_{(1)}^{*}\}&=&2|T_{(1)}|^2 = M + |z_1|  \geq 0 \ ,
\nonumber\\
\{T_{(2)} , T_{(2)}^{*}\}&=&2|T_{(2)}|^2 = M + |z_2|  \geq 0 \ .
\label{z}
\end{eqnarray}
The positivity bound for $M-|z_1|$ and $M-|z_2|$ exists because these
combinations of the mass and central charges of some state can be
expressed
through the square of particular supersymmetry generators acting on
that
state.

The first bound is saturated, i.e. $M =| z_1| $, if and only if the
state is
invariant
under one quarter of all the supersymmetries, since
 the state has to be invariant under the action
of $S_1, S_1^{*}$. The saturation of the second bound
$M =| z_2| $ means that the state has to be invariant under another
quarter of
the supersymmetries, $S_2, S_2^{*}$. Thus, if both bounds are
saturated,
i.e. $M =| z_1| =| z_2| $, the state has to be invariant under half
of all
supersymmetries.
For globally supersymmetric Yang-Mills theory, these bounds
are known as Bogomolny bounds for magnetic monopoles. To identify the
central charges one can quantize the theory, construct the
supersymmetry charges in terms of coordinates and canonical
momenta, and calculate the commutators of supersymmetry charges,
paying attention to boundary terms as was done in \cite{WO}
for $N=2$ globally supersymmetric Yang-Mills theory.

The situation with supersymmetric positivity bounds for
theories with local supersymmetries including gravity is in general
much more complicated. First of all, one can apply the bound only
for configurations which are asymptotically flat by identifying
the mass as the ADM or Bondi mass. The positivity of energy in
Einstein theory was obtained via supergravity theory by Deser,
Teitelboim and Grisaru \cite{DT}. Using the supergravity-type
formalism, Witten has presented a proof of the positivity
bound for the ADM or Bondi mass  of an asymptotically flat space
under the assumption of the dominant energy condition. This was
developed later to the so-called  Witten-Nester-Israel construction
\cite{W}.
 It has been shown in
\cite{W} that the mass of an asymptotically flat space-time is
non-negative
and vanishes only when the space-time is flat.
In terms of the Schwarzschild black hole  we may interpret the $N=1$
supersymmetry bound $M \geq 0$  as the statement that the $r=0$
singularity
is inside the horizon, until the bound is saturated, i.e the mass
vanishes.
However, when the mass vanishes the space-time becomes trivial.
The Schwarzschild black hole does not admit any unbroken
supersymmetry;
but broken supersymmetry does work as a cosmic censor as it requires
$M>0$.

The positivity bound for $N=2$ theory was derived in \cite{GH},
 by applying the  Witten-Nester-Israel construction \cite{W} to the
local $N=2$ supersymmetry transformation rules of the gravitino.
For asymptotically flat solutions of $N=2$ supergravity the
corresponding
bound is \cite{GH}
\begin{equation}\label{cen}
M\geq \sqrt {Q^2 + P^2}= |z_1|\ ,
\end{equation}
where the central charge $z_1$ has been expressed through the
electric and
magnetic charges.

For global $N=2$ supersymmetry of asymptotic states there is one
central
charge
 $z_1$ in the positivity bound, according to \cite{WO,FSZ}. Its
value,
\begin{equation}
 z_1 = \sqrt {Q^2 + P^2} \ ,
\end{equation}
has also been identified by Gibbons and Hull by solving the equations
\begin{equation}\label{sp}
\d \Psi_{\m I}  = \Bigl(\hat \nabla_\m (g , A )\,  \e \Bigr)_I =
\nabla_\m (g
)\e_I - (2\sqrt{2})^{-1}
\s^{\r\s}F^+_{\r\s}\epsilon _{IJ} \g_\m  \e^J
  \ , \quad I = 1, 2 \ \ ,\label{spin}
\end{equation}
 where the supercovariant connection in equation
(\ref{spin}) depends on the metric and the vector field (our
notation is defined in Appendix~A).
They have found that extreme Reissner-Nordstr\"om
 black holes with $M = \sqrt {Q^2 + P^2}$ admit super-covariantly
constant spinors of $N=2,\ d=4$ supergravity (\ref{sp}) and saturate
the
bound (\ref{cen}). The bound is saturated only
 for extreme  Reissner-Nordstr\"om black holes, as a consequence of
the
  fact that they admit supercovariantly
constant spinors, defined in eqs. (\ref{sp}).  These equations, in
the case
that
they have solutions, define the unbroken part $(N=1)$ of the
original $N=2$ supersymmetry of the theory.   {\it This shows that
the
solution admits some supersymmetries despite being purely bosonic;
the
generators of unbroken supersymmetry leave fermions invariant}
\footnote{The supersymmetry variation of the bosons contains only
fermions
and so is zero trivially for a bosonic solution.}.

 The positivity bound for Einstein-Maxwell theory, derived from
$N=2$ supergravity in \cite{GH} is in fact different from the
Bogomolny
bound for monopoles, as discussed in \cite{GG}. In general, {\it  the
identification of the central charges in the supersymmetry algebra,
which is necessary to derive the positivity bound (\ref{z}), is not
universal}.
 It depends both on the dynamics
and on the properties of the solutions.

To see the relation between the $N=2$ supersymmetry bound and the
cosmic censorship conjecture,  consider the charged
Reissner-Nordstr\"om
black hole with electric and magnetic charges $P$ and $Q$.  The
quantity $q$
appearing in the metric (\ref{2}) is $q = \sqrt {Q^2 + P^2}$.
In order that the singularity at $r=0$ be hidden by an event horizon,
we have
to require that
\begin{equation}
M\geq q \ ,
\end{equation}
just as in the purely electric case considered in the beginning of
this
section. This condition coincides with the
 requirement following from supersymmetry \cite{GH}.  Therefore, from
the
point of view of $N=2$ supersymmetry there exist special solutions of
the
Einstein-Maxwell equations, which happen to coincide with extreme
Reissner-Nordstr\"om  black holes and which solve not only the second
order
Einstein-Maxwell differential equations, but also the {\it first
order
differential equations} for spinors (\ref{sp}). This does not happen
for
non-extreme charged black holes.

In this paper we will investigate the corresponding issues for the
 dilaton black holes \cite{G} - \cite{HW}.

The positivity bound (\ref{z}) for an asymptotically
flat space consists of two equations:
\begin{eqnarray}\label{bb}
M &\geq & |z_1| \ ,\nonumber\\
M &\geq & |z_2| \ ,
\end{eqnarray}
since for $N=4$ there are two central charges, according to
\cite{FSZ}.
We will identify
those two central charges by considering the local supersymmetry
algebra.
It will also be shown that extreme black holes saturate
the supersymmetry bound  (either one of them for the solutions with
electric
and magnetic charge or both for the solutions with only electric or
only
magnetic charge).

The fundamental {\it first order differential equations} of $N=4$
theory,
which will be solved to produce extreme dilaton black holes,
generalize
those of $N=2$ theory, given in Eq. (\ref{sp}).
The four gravitinos and four dilatinos  will be required to have
vanishing local supersymmetry transformations in presence of gravity
$g_{\m\n}$, dilaton $\phi$, electric $A_\m$ and magnetic $B_\n$
fields.
\begin{eqnarray}\label{spi}
&&\d \Psi_{\m I}  = (\hat \nabla_\m (\phi, g , A, B )\,  \e )_I = 0 \
,
\nonumber\\
&&\nonumber\\
&&\d \L_I = - ( \hat {\dsl } \phi  (\phi, g , A, B ) \, \e )_I = 0\ ,
  \,  \hskip 1 cm  I = 1, 2, 3, 4.
\end{eqnarray}

\section{ Dilaton black holes}
 A dimensionally reduced superstring theory in $d=4$ can be
described in terms of $N=4$ supergravity. The latter exists in two
versions. One usually refers to the original one as the $SO(4)$
version
\cite{N4SO4}, and to the second one as the $SU(4)$ version
\cite{CSF}.
For the latter
the action is invariant under a rigid $SU(4)\otimes SU(1,1)$
symmetry, which makes that theory simpler.  In both versions, the
vector fields transform under an $SO(4)\equiv SU(2)\otimes SU(2)$
group. In the $SO(4)$ version both factors contain 3
vector fields, while in the $SU(4)$ version one factor consists of
three vector fields  and the other has three axial-vector fields. We
will consider $U(1) \otimes U(1)$ solutions, i.e.  solutions with one
non-trivial vector in each subgroup. There is also a complex scalar.
Its real
part is the dilaton  and its imaginary part is the axion.  We will
look for
solutions
which depend only on the  dilaton $\phi$; the axion field will be put
to a constant.
The remaining bosonic part of the action in these two cases is given
by
\footnote{The parameter
$a$ (or $g$), which governs the strength of the coupling of the
dilaton to
vector fields we keep always equal to 1, as
required by $N=4,\ d=4$ supersymmetry, and as in superstring theory,
i.e.  we consider only the case $a=1$ which has been qualified in
\cite{HW} as enigmatic.  However, for the $a=1$ case, the difference
with other investigations of charged dilaton black holes
\cite{GHS,HW}
 is the presence of two charges, electric and magnetic,
simultaneously,
 in the absence of the axion.}
 \begin{eqnarray}
 I_{SO(4)}& =&\int d^4x\,\sqrt{-g}\left( -R +2\partial^\mu
\phi\cdot\partial_\mu \phi
-\left( e^{-2\phi}F_{\mu\nu}F^{\mu\nu}
+e^{2\phi}\tilde G_{\mu\nu}\tilde G^{\mu\nu}\right)\right) \
,\nonumber\\
 I_{SU(4)}& =&\int d^4x\,\sqrt{-g}\left( -R +2\partial^\mu
\phi\cdot\partial_\mu \phi
-e^{-2\phi}\left(F_{\mu\nu}F^{\mu\nu}
+G_{\mu\nu}G^{\mu\nu}\right)\right)\ ,\label{actSU4N4}
\end{eqnarray}
 where
\begin{eqnarray}
F_{\m\n} &=&\partial_\m A_\n - \partial_\n A_\m \ ,\nonumber\\
\tilde G_{\mu\nu}&=&\partial_\mu \tilde B_\nu -\partial_\nu \tilde
B_\mu \ ,\nonumber\\
G_{\m\n} &=&\partial_\m B_\n - \partial_\n B_\m\ .
\end{eqnarray}
The actions are almost the same, except from the terms depending on
the vector $B$ or $\tilde B$.  The equations of motion of the two
theories are equivalent.  In fact, those of the $SO(4)$ version are
\begin{eqnarray}
\nabla_{\m}(e^{-2\phi}F^{\mu\nu})&=&0 \ ,\nonumber\\
\nabla_{\m}(e^{2\phi}\tilde
G^{\mu\nu})&=&0 \ ,\nonumber\\ \nabla^{2}\phi - \half e^{-2\phi}F^{2}
+
\half e^{2\phi}\tilde G^{2}& = &0 \ ,
\nonumber\\
R_{\m\n}  +2
\nabla_{\m}\phi\cdot \nabla_{\n}\phi-
e^{-2\phi}(2F_{\m\l}F_{\n\d}g^{\l\d}-\half
g_{\m\n}F^{2})&&\nonumber\\  - e^{2\phi}(2\tilde G_{\m\l}\tilde
G_{\n\d}g^{\l\d}-\half g_{\m\n}\tilde G^{2}) & = &0\ , \label{mot2}
\end{eqnarray}
while from the $SU(4)$ version we obtain
\begin{eqnarray}
\nabla_{\m}(e^{-2\phi}F^{\m\n})&=&0 \ ,\nonumber\\
\nabla_{\m}(e^{-2\phi}G^{\m\n})&=&0 \ ,\nonumber\\
\nabla^{2}\phi - \half e^{-2\phi}F^{2}  -
 \half e^{-2\phi}G^{2}& = &0 \ ,
\nonumber\\
R_{\m\n}
+2 \nabla_{\m}\phi \cdot\nabla_{\n}\phi-
e^{-2\phi}(2F_{\m\l}F_{\n\d}g^{\l\d}-\half
g_{\m\n}F^{2})&&\nonumber\\
- e^{-2\phi}(2G_{\m\l}G_{\n\d}g^{\l\d}-\half g_{\m\n}G^{2}) & =
&0 \ .\label{mot1}  \end{eqnarray}
The equivalence of these equations \cite{CSF} can be demonstrated
through the
duality rotation \begin{equation}\label{dual}
\tilde G^{\m\n} = \half i(-g)^{-\half}e^{-2\phi}\e^{\m\n\l\d}
G_{\l\d}\ .
\end{equation}
Both $G$ and $\tilde G$ are real with our conventions.
Such a duality rotation \cite{dual} transforms the
equation of motion of $\tilde B$ (the second line) to the Bianchi
identity of $B$, while the field equation of $B$ is the Bianchi
identity of $\tilde B$.  The other field equations are mapped into
each other.  Note that this transformation does not transform one
action in the other, a minus sign difference occurs for the $G$ and
$\tilde
G$ terms in  a space-time of Lorentzian signature \footnote{In
Euclidean
signature the two actions are connected by duality, though.}.

The solution of this system of equations has been given by Gibbons
\cite{GG},
and discussed in detail later by Gibbons and Maeda \cite{GM}. For
constant
axion, each vector field $A_{\m}$ and $B_{\m}$ (or $\tilde B_\mu $)
has to
be either electric
or magnetic, to satisfy the axion field equation, which reduces to
\begin{equation}
F_{\mu\nu}{} ^{\star} F^{\mu\nu} + G_{\mu\nu}{} ^{\star} G^{\mu\nu} =
0\ .
\end{equation}
The purely magnetic (electric) dilaton black holes have been studied
in
\cite{GHS,HW}.
Our solution generalizes the one
given in \cite{GG} by including asymptotically nonvanishing dilaton
field $\phi_0$ and the ones given in \cite{GHS},
\cite{HW} by keeping  both electric and magnetic charge. We will in
fact take $A_\mu $ to be purely electric, and $B_\mu $ to be
magnetic.
This implies that $\tilde B_\mu $ is also electric, and the
calculations are often simpler when using the electric solution
$\tilde B$, rather than the magnetic $B$.

Our solution depends on four independent parameters: $M, Q, P,
\phi_0$.
The mass of a black hole is $M$, the asymptotic value of the dilaton
field is $\phi_0$.
The electric charge of the $F$-field  is $ Q_{elec} = e^{\phi_0}Q$
and
the magnetic
charge of the $G$-field  $ P_{magn} = e^{\phi_0}P$, or equivalently,
the electric charge of the $\tilde G$-field is  $ P_{elec} =
e^{-\phi_0}P$. \footnote{Only when the asymptotic value of the
dilaton is zero does the electric charge equal $Q$ and the magnetic
one
equal $P$. We
have chosen the definition of charges in presence of  $\phi_0$ in a
way which
simplifies equations, since it is the parameters $P , Q$  which
appear
in all equations rather than $ Q_{elec}, P_{magn} $. }

There are few combinations of these parameters which will appear
in the solutions.

1. The dilaton charge, which is not an independent variable, is given
by \cite{GHS}
\begin{equation}
 \Sigma =   \frac {P^2 - Q^2 }{2M}\ ,
\end{equation}
where $ \Sigma$ is defined by the equation $\phi \sim  \phi_0 +
\Sigma/r$ at $r
\rightarrow \infty$.

2. The parameter $r_0$, which vanishes when the black hole becomes
extremal, is given by

\begin{equation}
r_{0}^2 = M ^2 + \S ^2 - P^2 - Q^2 \ \, = \ \, M ^2 + \S ^2 - e^{-2
\phi_0}P_{magn}^2 - e^{-2 \phi_0}Q_{elec}^2\ .
\end{equation}

 3. The outer and the inner horizons are defined in terms of a mass
and $r_0$,
\begin{equation}
r_{\pm} = M \pm r_0 \ .
\end{equation}
The solution of equations  (\ref{mot2}) can be given in the following
form.
\begin{eqnarray}\label{sol2}
&&ds^2=
e^{2U} dt^2 -
e^{-2U}
dr^2 -R^2  d\Omega  \ ,\nonumber\\
&&\nonumber\\
&&e^{2 \phi} =e^{2 \phi_0}\  \frac{r+ \Sigma}{r  - \Sigma} \ ,
\nonumber\\
&&\nonumber\\
&& F = \frac{Q\, e^{\phi_0}}{(r - \S)^2}\  dt \wedge dr \ ,
\nonumber\\
&&\nonumber\\
&& \tilde G = \frac{P\, e^{-\phi_0}}{(r + \S)^2} \  dt \wedge dr \ ,
\end{eqnarray}
where
\begin{equation}
e^{2U} =  \frac{(r - r_+) (r -  r_- )}{R^2}
\end{equation}
and
\begin{equation}
R^2 = r^2  - \Sigma^2 \ .
\end{equation}
The curvature singularity occurs at $r = |\Sigma|$.

The solution has manifest dual symmetry:
\begin{eqnarray}
Q&\leftrightarrow & P \ , \nonumber\\
\Sigma&\leftrightarrow & - \Sigma \ ,\nonumber\\
F&\leftrightarrow & \tilde G \ ,\nonumber\\
\phi&\leftrightarrow &- \phi\ .
\end{eqnarray}

To write our solution in a form which corresponds to solution of eqs.
(\ref{mot1}), we have to add to eqs. (\ref{sol2}) the result for the
non-dually rotated field $G_{\m\n}$.
\begin{equation}\label{sol1}
G = P\, e^{\phi_0}\sin \t \, d\t \wedge d\phi =  P_{magn}\,  \sin \t
\,d\t
\wedge d\phi\ .
\end{equation}

Notice that our solution also yields a solution of a theory with a
smaller
field content $(g_{\mu\nu}, \phi, \cal F_{\mu\nu})$ and action

\begin{equation}
I_{\cal F} = \int d^4x\,\sqrt{-g}\left( -R +2\partial^\mu
\phi\cdot\partial_\mu \phi
-e^{-2\phi} {\cal F}_{\mu\nu} {\cal F}^{\mu\nu} \right).
\end{equation}
To see this, take $\cal F$ to be both electric and magnetic,
\begin{equation}
{\cal F} = F + G = \frac{Q\, e^{\phi_{0}} }{ (r - \Sigma)^2}\  dt
\wedge dr +
                P \, e^{\phi_{0}}\, sin \theta \, d\theta \wedge
d\phi \ ,
\end{equation}
and note that the energy-momentum tensor of $ \cal F$ is just that of
the two fields $F$ and $G$, because the electric and magnetic fields
are parallel.  This implies that the equations of motion for this
theory are
consistent with those of the original one. All thermodynamic
properties, which
are controlled by the metric, will be indifferent to whether we use
this
arrangement of fields or our original one.

Let us also introduce the following notation:
\begin{equation}
z_1 =\frac  {Q - P}{\sqrt 2 }  \ ,\qquad
z_2 =\frac  {Q + P}{\sqrt 2 } \ ,
\end{equation}
so that
\begin{equation}
Q_{elec} = e^{\phi_0}\ \frac {z_1 + z_2 } {\sqrt 2 } \ ,\qquad
P_{magn} = e^{\phi_0}\  \frac {z_2 - z_1 } {\sqrt 2 }\ .
\end{equation}
Later on, we will identify these combinations of electric and
magnetic
charges as central charges in the supersymmetry algebra.
The dilaton charge in this notation is given by
\begin{eqnarray}
\Sigma = - \frac{z_1z_2}{M} \ ,
\end{eqnarray}
and the parameter $r_0$ showing the deviation from extremality is
\begin{equation}
r_{0}^2 = {1\over M^2}\ (M^2 - z_1^2)  (M^2 - z_2^2)  \ .
\end{equation}

In Sec.  6  the supersymmetric properties of the dilaton black holes
will
be studied and it will be shown that supersymmetry leads to
the positivity bound (\ref{b}), which implies that
\begin{eqnarray}\label{CS}
M &\geq & |z_1| \ , \nonumber\\
M &\geq & |z_2 |\ .
\end{eqnarray}
Either of these inequalities can be saturated only if at least $N=1$
supersymmetry is unbroken, see Sec.  6. In this case $r_0$ vanishes
and we
deal with extreme black holes.

Eq. (\ref{CS}) implies that the  parameters
of the  dilaton black hole can vary only inside the square
\begin{equation}
 {|Q|+|P|\over \sqrt 2} \equiv M_{extr}  \leq \, M \,  \  .
\end{equation}
It is instructive to consider various  special cases of the dilaton
black hole
(\ref{sol2}) for a given mass $M$, see Fig. 1.

\begin{enumerate}
\item The Schwarzschild solution is given by eqs. (\ref{sol2}) at

$P=Q=\Sigma= \phi_0 = 0, \, r_+=2M, \, r_- = 0$ .  This solution
corresponds
to the point at the centre of coordinates in Fig.  1.
\item Classical Reissner-Nordstr\"om black hole with equal electric
and magnetic charges. $|P| = |Q|$, \, $\Sigma = \phi_0 = 0$. This
solution corresponds to the
 lines crossing the centre of coordinates which are parallel to the
boundaries of the square in Fig. 1.

\item Purely magnetic dilaton black hole described in
\cite{GHS}.
$Q=0$, \, $-z_1 =z_2 =  P/{\sqrt 2} $, \, $\Sigma=P^2/{2M} $
 and $r_0 = M - \Sigma ,  \quad r_- = \Sigma , \quad r_+ = 2M -\Sigma
$.
By performing the change of variables $r' = r + \Sigma$,
we recover the metric as given in \cite{GHS}. This solution
corresponds to the P-axis in Fig. 1.

\item Purely electric dilaton black hole described in
\cite{HW}. Change $Q$ to $P$, \, $\Sigma
$ to $- \Sigma$ in the previous case. The solution corresponds to the
Q-axis in Fig. 1.

\item Extreme black holes with  electric and magnetic charges. $M =(
|Q| +
|P|)/(\sqrt 2 ) $,\, $\Sigma
= (|P| - |Q|)/(\sqrt 2 )$, $M> |\S|$ \, $r_0 = 0$, $r_+ = r_-  = M$.

For $PQ > 0$, \ $M = |z_2| > |z_1|$, while
for  $PQ < 0$, \ $M = |z_1| > |z_2|$. These solutions will be
discussed later.
In
Fig. 1 they correspond to the boundary of the square excluding the
four
vertices .

\item Extreme black holes with either  electric or  magnetic charge.
$r_+ = r_-  = M$, \, $r_0 = 0$, \, $M=|z_1|= |z_2| = |\Sigma |$. In
Fig. 1
these solutions  correspond to the four vertices of the
square and are the stringy extreme charged electric or magnetic
dilaton
black holes of \cite{GHS,HW}.
 \end{enumerate}

There are many ways to generalize the black hole solutions which we
have
presented above. We considered a simple solution
depending on two charges, $P$ and $Q$, i.e. we assumed that each of
the fields $F$ and $G$ has either electric or magnetic charge, but
not both. However,  these solutions  can be easily generalized to
solutions in
which both fields  $F$ and $G$  have electric and magnetic charges.
The
general form of these solutions is the same as that of our solutions,
with
$P^{2}_{F}+P^{2}_{G}$ replacing $P^{2}$ in all our equations,  and
similarly
for $Q$. (Only products of two $F$s or two $G$s appear in the
equations of motion.) These solutions are consistent with a constant
axion
if $Q_{F}P_{F}+Q_{G}P_{G}=0$. Thus, instead of solutions
characterized by
two parameters, $P$ and $Q$, we essentially have a set of solutions
depending on three independent parameters. All the properties that
hold for
the solutions we have studied and depend only on the metric, remain
true for
the new set of solutions.

\section{Extreme dilaton black holes}
We will look for a static solution of equations of motions of the
theory
(\ref{actSU4N4}),
not necessarily spherically symmetric, with the following ansatz for
the metric in isotropic coordinates (conformastatic metric):
\begin{eqnarray}
ds^2 &=& e^{2U} dt^2 - e^{-2U}(dx^i)^2\ .
\label{genmet}\end{eqnarray}
The nonzero components of the vierbein $e_\mu{}^a$ are
\begin{equation}
e_{\hat 0}{}^0=e^U\ ,\qquad e_{\hat i}{}^j= e^{-U}\delta_i{}^j\ ,
\end{equation}
where $g_{\mu\nu}=e_\mu{}^a e_\nu{}^b \eta_{ab}$, and the function
$U$  is time-independent. For the determinant of the metric
we have $\sqrt{-g}=e^{-2U}$. The non-zero elements of the
spin-connection $\omega_\mu{}^{ab}$ and Ricci tensor $R_{\mu\nu}$ are
given by
 \begin{eqnarray}
\omega_0{}^{i0}&=&e^{2U}\partial_i U \ ,\nonumber\\
\omega_i{}^{jk}&=&2\delta_{i[j}\partial_{k]}U \ ,\nonumber\\
R_{00}&=&-e^{4U}\partial_i\partial_i U \ ,\nonumber\\
R_{ij}&=&2\partial_i U\cdot \partial_j U-\delta_{ij}
\partial_k\partial_k
U \ .
 \end{eqnarray}
We look for solutions where $A_\mu$ is electric and $B_\mu$ is
magnetic (or
$\tilde B_\mu$ is again electric),
\begin{equation}
A_\mu=\delta_\mu^0 \psi\ ,\qquad\tilde B _\mu=\delta_\mu^0 \chi\ .
\end{equation}
For $G_{\mu\nu}$ this implies that it has only space-like components
given by
\begin{equation}
\half \epsilon^{ijk} G_{jk}=-e^{2\phi-2U}\partial_i \chi\ .
\end{equation}

Then the field
equations \eqn{mot2} in this metric are
\begin{eqnarray} \partial_i e^{-2U-2\phi} \partial_i \psi
&=&0 \ ,\nonumber\\
 \partial_i e^{-2U+2\phi} \partial_i \chi &=&0 \ ,\nonumber\\
- \partial_i  \partial_i \phi + e^{-2U-2\phi}(
\partial_i\psi)^2-e^{-2U+2\phi}(
\partial_i \chi)^2
&=&0 \ ,\nonumber\\
- \partial_i \partial_i U+e^{-2U-2\phi}(
\partial_i\psi)^2+e^{-2U+2\phi}(
\partial_i
\chi)^2&=&0 \ ,\nonumber\\
 \partial_i U\cdot  \partial_j U+ \partial_i\phi \cdot  \partial_j
\phi&&\nonumber\\ -e^{-2U-2\phi}\partial _i\psi\cdot  \partial_j\psi
-e^{-2U+2\phi} \partial_i\chi\cdot  \partial_j\chi&=&0\ .
\end{eqnarray}
 We define
\begin{equation}\label{H1H2}
H_1=e^{-U-\phi} \ , \qquad H_2=e^{-U+\phi}\ .
\end{equation}

These equations can be solved as
follows:

  \begin{eqnarray}
ds^{2} &=&  e^{2U}dt^{2}-e^{-2U}d\vec{x}^{2} \ ,\nonumber\\
A = \psi dt  &, &\qquad \tilde B = \chi dt  \ , \nonumber\\
F =  d \psi \wedge dt &, &\qquad  \tilde G = d \chi \wedge dt \ ,
\nonumber\\
e^{-2U} = H_1 H_2  &, &\qquad
e^{2\phi} = H_2/ H_1  \ ,\nonumber\\
\sqrt{2}\, \psi = \pm H_1^{-1} &,&\qquad  \sqrt{2}\,\chi
=\pm H_2^{-1} \ ,\nonumber\\
\partial_i\partial_iH_1 =0 &,&\qquad \partial_i\partial_i H_2=0\ .
\label{extsol}
\end{eqnarray}

Thus, two  arbitrary harmonic functions $H_1, H_2 $ can be used to
build
the metric,  dilaton and vector fields according to eqs.
(\ref{extsol}).
Specific examples are given below.

 i) The {\it extreme multi black hole} solution is the solution  of
the
equations given above with \begin{eqnarray}\label{nholes}
H_1 &=& e^{-\phi_0} (1+ \sum _{s=1}^{n}\frac{\sqrt {2
}|Q_s|}{|x-x_s|})  \ ,
\nonumber\\ H_2 &=& e^{+\phi_0} (1 + \sum _{s=1}^{n}\frac{\sqrt {2
}|P_s|}{|x-x_s|})\ , \end{eqnarray}
where there is a following relation between the parameters of each
black
hole.
\begin{eqnarray}
M_s &=&\frac{ |P_s| + |Q_s|}{\sqrt {2 }} \ ,\nonumber\\
\nonumber\\
\S_s &=&\frac{ |P_s| - |Q_s|}{\sqrt {2 }}  .
\label{multi}  \end{eqnarray}
It follows that
\begin{equation}\label{bs}
M_s^2 +\S_s^2 = P_s^2 + Q_s^2 \ .
\end{equation}
This allows a static equilibrium due to the balance of gravitational,
scalar
and electromagnetic forces.
The total mass and charges of the full configuration are given by
\begin{eqnarray}
M = \sum_{s=1}^{n} M_s  \ , &\qquad &
\S = \sum_{s=1}^{n} \S_s  \ ,\nonumber\\
P = \pm \sum_{s=1}^{n} |P_s|  \ ,&\qquad&
Q = \pm \sum_{s=1}^{n} |Q_s| \ ,\nonumber\\
sgn(P) = sgn(P_s)  \ , &\qquad &
sgn(Q) =  sgn(Q_s) \ .
\label{total}  \end{eqnarray}
They also satisfy the condition
\begin{equation}
M^2 +\S^2 = P^2 + Q^2 \ .
\end{equation}

To see the force balance explicitly, let us consider Newtonian,
Coulomb and
dilatonic forces.  The force between two distant objects of masses
and charges
$(M_1, Q_1, P_1, \S_1)$ and $(M_2, Q_2, P_2,\S_2)$ is
\begin{equation}\label{bal}
F_{12} = - \frac{M_1 M_2}{r_{12}^2} + \frac{Q_1 Q_2}{r_{12}^2} +
\frac{P_1 P_2}{r_{12}^2} - \frac{\S_1 \S_2}{r_{12}^2} \ .
\end{equation}
The dilatonic force is attractive for charges of the same sign and
repulsive
for
charges of opposite sign.
Using the relations (\ref{multi}) for the masses and dilaton charges
 in terms of the magnetic and electric charges, we
see  that $F_{12}$ vanishes.  In particular, it follows that a purely
magnetic and
a purely electric extreme black hole can be in equilibrium, as the
attractive
gravitational force is balanced by the repulsive dilatonic force.

  The extreme electrically (or magnetically) charged multi black hole
solutions
are solutions of the type (\ref{nholes})
with $P_s =0$ (or $Q_s =0$).
They can be formally identified with a special
case of a metric of Bonnor \cite{B} describing charged dust in
equilibrium. \footnote{The relevance of Bonnor metrics with charged
dust in
equilibrium
to metrics admitting super-covariant constant spinors in the context
of $N=2$ supergravity was discovered by Tod \cite{T}.}
The corresponding equations are
\begin{eqnarray}
\nabla_{\m} F^{\m\n}&=& J^\n  \ ,\nonumber\\
  R_{\m\n} -(2F_{\m\l}F_{\n\d}g^{\l\d}-\half g_{\m\n}F^{2})
& = & T^{\m\n}\ ,
\label{bon}  \end{eqnarray}
where
\begin{eqnarray}
T^{\m\n} &=& \varepsilon u^\m u^\n  \ ,\nonumber\\
J^\m &=& \s u^\m \ ,
  \end{eqnarray}
and $u^\m$ is the four-velocity of dust with
normalization
$g_{\m\n} u^\m u^\n =1$.
The charged dust in equilibrium is
characterized,
according to Bonnor \cite{B} by the condition
$\varepsilon =  \pm \s$.
We have found that if the density of charged dust in equilibrium is
\begin{equation}\label{eps}
\varepsilon (x)=  \pm \s(x) = - 2 (\nabla \phi )^2 (x)\ ,
\end{equation}
or equivalently, the trace of the energy momentum tensor of dust is
proportional
to the
scalar curvature of the space due to the presence of the dilaton
\begin{equation}
   T = R =  - 2 (\nabla \phi )^2 \ ,
\end{equation}
then the Bonnor solution of the system of equations (\ref{bon}) -
(\ref{eps})
coincides with the set of
extreme electrically (or magnetically, after duality transformation)
charged dilaton black holes.
The reason behind the formal identification of extreme dilaton black
holes
with charged dust is the following~: The energy momentum tensor in
our eqs.
(\ref{mot2}), (\ref{mot1}) is covariant; however, on solutions it
coincides
with
the non-covariant energy momentum tensor of the charged dust in the
Bonnor
eq. (\ref{bon}).

\vskip  0.6 cm

 ii) As a specific example of extreme black holes, let us now
consider
an  extreme electrically  and
magnetically charged {\it spherically symmetric} black hole with
dilaton
field not vanishing at infinity:

\begin{equation}
ds^2 = \Bigl( 1 + \frac{\sqrt 2 ( | Q | +| P |)}{\r} +
\frac{2|PQ|}{\r^2}
\Bigr)
^{-1}  dt^2 - \Bigl(1 + \frac{\sqrt 2 ( | Q | +| P |)}{\r} +
\frac{2|PQ|}{\r^2}
\Bigr)  d \vec{x}^{2}
\end{equation}
where $\r = |\vec{x}|$ and
\begin{equation}
e^{2\phi} = e^{2\phi_0} \Bigl( \frac{\r + {\sqrt 2}  | P |}{\r +
{\sqrt 2}  | Q
| }
\Bigr) \ ,
\end{equation}
i.e., we choose
\begin{equation}\label{H}
H_1 = e^{-\phi_0} \Bigl(1 + \frac{\sqrt 2  | Q |}{\r} \Bigr) \
,\qquad
H_2 = e^{\phi_0} \Bigl(1 +
\frac{\sqrt 2   |P |}{\r} \Bigr)\ .
\end{equation}

This solution can be compared with the one described in the previous
section
(case 5 in the list)
under a suitable change of variables:
\begin{equation}
r = \r+ M \ .
\end{equation}
 The electrically  and magnetically charged
spherically symmetric dilaton black hole  (\ref{sol2}) with
\begin{eqnarray}
r_+ &=& r_-  = M \ ,\nonumber\\
r_0 &=&0 \ ,
\end{eqnarray}
 is the extreme dilaton black hole.
This is in accordance with Fig 1. where the boundary of the square
(excluding the vertices) corresponds to the extreme
black hole with both electric and magnetic charge.
The mass and charges of the   extreme dilaton black hole with
nonvanishing
electric and magnetic charge  satisfy the bounds
$M =\frac{ | P| +  |Q |}{ \sqrt 2}\ ,\quad
M > \Sigma = \frac{ | P| -  |Q |}{ \sqrt 2}$ .
Thus for the generic extreme black hole with electric and magnetic
charges only one of the
the positivity bounds (\ref{CS}) required for cosmic censorship is
saturated.
For $PQ > 0$ (sides I and III of the square in Fig. 1)
$M = |z_2| $
and the second one is still a positivity bound since
$M  >   |\S |= |z_1|$.
For sides II and IV  with $PQ<0$ we have
$M = |z_1|$ ,
and
$M  >   |\S |= |z_2|$.
Thus, all over the boundary
of the square in Fig. 1, except for the vertices, the absolute value
of the
dilaton
charge is smaller than the mass.
This property of the extreme
black hole with both electric and magnetic charge means that the
singularity
$r =  |\S| $ is inside the horizon $r_+ = r_- = M$.
The purely magnetic (electric) extreme black holes are the solutions
given in Eq.
(\ref{H}) with $P=0$ ($Q=0$).
 These solutions in Fig. 1 correspond to the four vertices of the
square.
 For these solutions
$M = |z_1| = | z_2| =  |\S| $ .
For purely magnetic extreme black holes, the singularity at $r = |\S|
=M$
coincides with the horizon $r_+ = r_- = M$. It
is argued, however, that for the metric which the string sees,
$ds^2_{str}  =
e^{2\phi} ds^2$, the horizon moves infinitely far away and the
curvature tensor
becomes nonsingular \cite{GHS}. For purely electric
extreme black holes this kind of argument
is absent.

It is important that for all these solutions the positivity bounds
(\ref{CS})
imply that the singularity $r =  |\S| $ is either inside the horizon
$r_+$ or
coincides with it. One can easily check that $r_+$ is, indeed, an
event
horizon, i.e. the integral in (\ref{3}) diverges for $r_1 = r_+$.
This happens
independently of our choice of normal metric versus  stringy one.
This means
that, in agreement with the cosmic censorship conjecture,
supersymmetry
saves an outside observer from seeing the singularity.

\section{Thermal properties of the dilaton black hole}

The explicit expression for the metric of the charged dilaton black
hole allows us to calculate its thermal properties. Detailed
analysis of such properties has been performed  in
\cite{GHS,HW} for the electric or magnetic stringy black holes
(the solutions on the $Q$- and $P$-axes of Fig. 1) or for non-stringy
black
holes with $e^{-2a\phi}$, $a\neq1$ in the action.  Discussion of some
of
the thermal  properties and singularities of electrically and
magnetically
charged black holes can be found in \cite{GM}.

In calculating the temperature and entropy of black holes, we must
keep in
mind that the interpretation of the results as {\it physical}
temperature and
entropy may not be reliable in some limits. In fact,  the thermal
description
of
purely electric or magnetic dilaton black holes breaks down near
extremality
\cite{HW}. At the end of this section we will analyze the breakdown
of the
thermal description for our class of charged dilaton black holes.
However, purely geometric quantities such as the area $A$
and the surface gravity $\kappa$ of the black hole horizon always
make sense
and the``thermodynamic" relationship between them will be seen to
hold  in
any case when a black hole has a non-singular horizon.

The Hawking temperature of the black hole (\ref{sol2}) can
be
calculated by a variety of standard methods. In terms of the surface
gravity
$\kappa$, it is given by $ T=\frac{\kappa}{2\pi}
$.  The surface gravity can be calculated from the Killing vector
$\zeta^{\mu}$,
which for any static metric
\begin{equation}
ds^{2}=g_{tt}(x)\,dt^{2}-h_{ij}(x)\,dx^{i}dx^{j}
\end{equation}
is simply
\begin{equation}
\zeta_{\mu}=\delta_{\mu t}\, g_{tt}\ .
\end{equation}
Thus, the surface gravity is
\begin{equation}\label{Surf}
\kappa   =   \Bigl[-\half(\nabla^{\mu}\zeta^{\nu})
(\nabla_{\mu}\zeta_{\nu}) \Bigr]^{\half}_{r=r_{horizon}}
= \frac{1}{2} \Bigl(\frac{dg_{tt}}{dr} \Bigr)_{r=r_{+}}
= \frac{1}{2} \  \frac{r_+ - r_-}{r_+^2 -  \S ^2} \ ,
\end{equation}
where again
\begin{equation}
\S = \frac{P^2 -Q^2}{2M}\ , \qquad
r_{\pm} = M \pm \sqrt {M^2 + \S^2 - P^2 - Q^2}\ .
\end{equation}

Then the  temperature of our black hole (\ref{sol2}) is given
by

\begin{equation}\label{Temp}
T   = \frac{1}{4\pi} \  \frac{r_+ - r_-}{r_+^2 -  \S ^2}\label{T} \ .
\end{equation}

The isothermals are drawn on the $P,Q$ plane in Fig. 2 for a fixed
value
of the mass $M$.
{}From Eq. (\ref{T}) a fixed-mass surface $T = T (z_1, z_2)$ is
plotted in Fig.
3.
The temperature falls very sharply to zero temperature near
extremality at
the borders of the regular black holes. In Fig. 4, the value of the
temperature
as a function of mass is plotted for different values of $P$ and $Q$.
By
inspection of the figures, we can see the following :
{\it  Extreme black holes with both  electric and magnetic charges
 have zero temperature. At the corners
there is a  discontinuity}. Consider as an example the
 purely magnetic extreme dilaton black hole. We may either first take
the limit
$Q\rightarrow 0$ in our expression for the temperature in Eq.
(\ref{Temp})
and after that take the limit to extreme $(r_+ - r_-) \rightarrow 0$,
or
vice-versa; the limiting temperature depends on which choice we make
:
\begin{equation}
\lim _{(r_+ - r_-) \rightarrow 0} \quad \lim _{Q\rightarrow 0}\,
T(P,Q,M)
 =\frac{1}{8\pi M}\ ,
\end{equation}
\begin{equation}
\lim _{Q\rightarrow 0} \quad \lim _{(r_+ - r_-) \rightarrow 0}
T(P,Q,M)
 = 0 \ .
\end{equation}
We can see this also from Figs. 3, 4.  Note that different results
can be
obtained by calculating the limit along different isothermals shown
in Fig. 2.
The limit can be anywhere between $0$ and $\frac{1}{8\pi M}$. The
fact that
the temperature at the corners
is not well-defined explains why there are apparently contradictory
statements about it in the literature \cite{GHS},
\cite{HW}.\footnote{One
possible indicator as to which temperature we should take is provided
by
supersymmetry, which is usually related to zero rather than finite
temperature.  We will show in Sec.  6 that the extreme purely
magnetic or
electric solutions possess two out of four possible supersymmetries,
so one
may prefer to take $T=0$.}

To investigate this problem further, let us examine carefully
the limits of applicability of the thermal description of black
holes.
There are several conditions which must be satisfied.
One of them was obtained and extensively discussed in Ref. \cite{HW}.
A thermal  description of a system is possible only if,  after an
emission of
a
single quantum of a typical energy $T$, the temperature of the system
changes
by  $|\Delta T| \ll T$. Applying this to a black hole of a mass $M$
gives,
according to  \cite{HW},
\begin{equation}\label{1cond}
\left|\left(\frac{\partial T}{\partial M}\right)_{P,Q}\right| \ll 1\
{}.
\end{equation}
This inequality may be rewritten as:
\begin{equation}\label{2cond}
T \left(\frac{\partial S}{\partial T}\right)_{P,Q}
\gg 1 \ .
\end{equation}
According to \cite{HW}, this condition has a profound
physical interpretation; it says that thermal description is possible
when the {\it available entropy} of the black hole, i.e. the number
of states
available     within its thermal energy interval, is very large.

The general belief expressed in   Ref. \cite{HW} was that thermal
description
breaks down for extreme black holes. However,  there remained some
confusion, since the criteria (\ref{1cond}) and (\ref{2cond}),
applied to
purely electric (or magnetic) black holes, did not show any signal of
danger
even arbitrarily close to the extreme point, when $M \rightarrow
Q/\sqrt 2$
(or $P/\sqrt 2$) \cite{HW}.  Our dilaton black holes correspond to
the
case $a = 1$ in terminology of Ref. \cite{HW}, where this case was
labelled
``enigmatic''. To clarify what is going on, let us go back to the
derivation of
(\ref{1cond}).

When a black hole emits a particle of typical energy T (from the
point of view
of
a static observer at infinity), its mass decreases,
$M \rightarrow (M~-~T)$.
Its
temperature $T(M)$ becomes $T((M~-~T))$ . The
condition $ \Delta T \ll T$  can be written as follows:
\begin{equation}\label{3cond}
 |T((M - T)) - T(M)| \ll T\ .
\end{equation}
If (and only if) the function $T(M)$ has a derivative which
remains almost constant in the interval between $M$ and $M - T$,  Eq.
(\ref{3cond}) can be rewritten in the form $T\, |\partial T/\partial
M| \ll T$,
which is equivalent to (\ref{1cond}).
In most theories studied in \cite{HW} all conditions
necessary for the derivation of (\ref{1cond}) are satisfied.
However,
for purely electric (or magnetic) dilaton black holes, we see
violation when
the
mass of the black hole approaches its extreme value $M_{extr} =
\frac{|P| +
|Q|}{\sqrt{2}}$.  For example, (\ref{3cond}) is violated near
extremality,
where any emission of  a quantum  of typical energy $T \approx
{1\over 8 \pi
M_{extr}}$ would reduce the mass below the
lower bound coming from supersymmetry, and this is absolutely
forbidden.
Thus, failure of the thermal description occurs for $PQ = 0$ holes
when
\begin{equation}\label{4cond}
\Delta M = M - M_{extr} \geq {1\over 8\pi M_{extr}} \ .
\end{equation}
To obtain further insight, let us approach this question in a more
general
context,
when both $Q$ and $P$ do not vanish.
Assume, for example, that $Q > P > 0$. In this case $z_2 > z_1 > 0$,
and the
black
hole becomes extreme when its mass decreases down to $M_{extr} = z_2
= (Q +
P)/\sqrt 2$. One can easily verify by using Eq. (\ref{Temp}), that
when the
mass
of the black hole approaches $z_2$, i.e. when $\Delta M = (M -
M_{extr})
\rightarrow 0$, the temperature of the black hole vanishes as
$\sqrt{\Delta
M}$,
\begin{equation}\label{xx}
T \sim {1\over 2 \pi}\left({2\ \Delta M\over z_2
(z_2^2 - z_1^2)}\right)^{1/2} \ .
\end{equation} Therefore, the expression for
$\left(\frac{\partial T}{\partial M}\right)_{P,Q}$ diverges in this
limit,
\begin{equation}\label{xxx}
\left(\frac{\partial T}{\partial M}\right)_{P,Q} \sim
{1\over 2 \pi}\, {1\over \sqrt{\Delta M} \sqrt{2 \, z_2 (z_2^2 -
z_1^2)}} \sim
\
{1\over 2 \pi Q}\, {1\over \sqrt{\Delta M } \sqrt{2\sqrt 2 \  P } }
\ .
\end{equation}
This result is illustrated by Fig. 4. Eq. (\ref{xxx}) implies that
the condition (\ref{1cond}) is always violated when the black hole
approaches
its
extreme limit. It is just more difficult to see it working in the
limit $P =
0$. We see, in particular, that in the limit $P \rightarrow 0$ the
thermal
description breaks down along the whole slope from $T_{extr} \sim
1/8\pi M$
to $T = 0$. Thus, the discrepancy between different ways of
calculating the
temperature of extreme electric (or magnetic) black holes is just one
manifestation of the breakdown of the thermal description in this
limit.

However, from our previous arguments it follows that thermal
description
breaks
down even earlier. Indeed, one can easily check that for small $P$
the
assumptions used in the derivation of Eq. (\ref{1cond}) break down,
and that
the thermal description becomes inapplicable even before the
temperature
reaches its maximum, just as in the case $P = 0$, see Eq.
(\ref{4cond}).

Additional information can be obtained by studying the behavior of
the
entropy.  The  expression in Eq. (\ref{2cond}) can be interpreted as
the
available entropy only if the temperature of the extreme black hole
is equal to
zero. Indeed, only in this case $\Delta T \equiv T - T_{extr} = T$
and one may
write that $\Delta S = (\partial S/\partial T) \, T$. As we noted
already,
the value of $T$ for extreme purely electric (or magnetic) black
holes is
ambiguous, due to
the failure of thermal description of  extreme black holes.
Fortunately, the
entropy can be calculated by several other methods, and the results
do not
depend on this ambiguity.

The entropy of a black hole can most easily be calculated as one
fourth of the area of the horizon $A$.  The physical
radial coordinate is $R$, so that the area of a sphere of
radius $R$ is simply $4\pi R^{2}$.  This gives
\begin{equation}\label{E}
S= \pi R^2 | ^{r=r_+}_{r=\Sigma } = \pi (r_+^2 - \S^2)\ .
\end{equation}

The thermodynamic relation
$T= \left(\frac{\partial S}{\partial M} \right)_{P,Q}^{-1}$ may be
readily
checked to be obeyed.  There is also a nice relation between the
temperature
and
the entropy of the charged dilaton black hole
\begin{equation}
S\, T =
\frac{1}{4}\, (r_+ - r_-)\ .
\end{equation}

One can check that Eq.  (\ref{E}) correctly describes all the
particular cases
listed in the end of Sec.  3, for which the entropy was already
known.
For example, it is easy to see from
(\ref{E}) that the entropy unambiguously vanishes at the corners of
our
square, when the horizon $r_+$ coincides with the value of the
dilaton
charge $|\S|$.  However, on the sides of the square (for $PQ \not =
0$ extremal
black holes) the entropy does not vanish.  Its value is
\begin{equation}
S_{extr} = \pi (M^2 - \S^2) = 2\pi |P Q|  = \pi \left|z_1^2 -
z_2^2\right|\ .
\end{equation}

All these properties can be seen in Fig. 5, where the
surface $S=S(z_1, z_2)$ (for fixed mass $M$) is plotted. Another
example of
extreme dilaton black holes with non-zero entropy has recently been
studied
by Horne and Horowitz \cite{HH}.  The angular momentum $J$ plays
there the
same role as the mixing of electric and magnetic charge plays here.

To calculate the {\it available } entropy $S_a$, one should subtract
$S_{extr}$ from  $S$, \begin{equation}\label{xxxx}
S_a=  S - S_{extr} = \pi (r_+^2 - \S^2) - 2\pi |P Q| \ .
\end{equation}
This quantity becomes much larger than 1 only far away from the
extreme
regime, as we can see by considering two particular cases:

i) If the black hole has only electric charge $Q > 0$, then we obtain
for the
regime $\Delta M \equiv M - M_{extr} \ll M_{extr}$
\begin{equation}\label{i2}
S =  S_a =  8\pi \ \Delta M\, M_{extr} \ ,
\end{equation}
while for $\Delta M  \gg M_{extr}$
\begin{equation}\label{i3}
S =  S_a =  4\pi \ (M^2 - {3\over 4} M^2_{extr}) \ .
\end{equation}

ii) If the black hole has $Q = P > 0$, then
in the case $\Delta M  \ll M_{extr}$ we have
\begin{equation}\label{ii3}
S_a =  2\sqrt 2 \pi \sqrt {\Delta M} \, M_{extr}^{3/2}\ .
\end{equation}
In the opposite limit, $\Delta M  \gg M_{extr}$,
\begin{equation}\label{ii4}
S =  S_a =  4\pi \ (M^2 - {1\over 2} M^2_{extr}) \ .
\end{equation}

The thermal description is valid only if
$S_a  \gg 1$. Note that the entropy $S = S_a$  of the purely electric
black
hole
is always {\it smaller} than the entropy of the black hole with both
electric
and
magnetic charge corresponding to the same $M_{extr}$, i.e. breakdown
of the
thermal description occurs earlier during the evaporation process for
$PQ=0$ holes. This result is rather surprising, since the
(uncritical) use of
Eq.
(\ref{1cond}) would have led us to the opposite conclusion.

To get a more quantitative estimate, let us assume that the thermal
description is applicable when $S_a >  N$, where $N$ is some large
constant,
$N \gg 1$. Let us assume also that the thermal description breaks
down (with
a decrease of $\Delta M$) in the domain where $\Delta M \ll
M_{extr}$. This
is possible only if $8\pi M^2_{extr}\gg N$.  In the purely electric
case the
thermal
description breaks down at
\begin{equation}\label{ii5} \Delta M <
{N\over 8\pi M_{extr}} \ .
\end{equation}
This condition is in an agreement with our earlier estimate
(\ref{4cond}), but
is even stronger. For large $M_{extr}$, the temperature of the
black hole at that time is very close to $(8\pi M_{extr})^{-1}$, but
it never
reaches this limit in the region where a thermal description is
possible.

In the case $P = Q$ the thermal description breaks down later,   at
\begin{equation}\label{ii6}
\Delta M < {N^2\over 8\pi^2 M^3_{extr}} \ .
\end{equation}
At this stage the temperature of the black hole is given by
\begin{equation}\label{ii7}
T \sim  {1\over 4 \pi^2}\ {N\over M^3_{extr}} \ .
\end{equation}
The larger the body,  the smoother the function $T(M)$, and the
smaller
the temperatures at which thermodynamics describes successfully the
black hole near its extremal state.

In Fig.  4 we also see that the temperature has a maximum for every
combination of the charges.  $\left( \frac{\pa T}{\pa M}
\right)_{P,Q}$
vanishes
there, has different sign at both sides, and  goes to zero when
we approach extremality. This means that the specific heat blows up
at the temperature maximum, changes sign there and goes smoothly to
zero
when we approach extremality. A direct calculation gives
\begin{equation}\label{C}
C^{-1} = \left(\frac{\partial T}{\partial
M}\right)_{P,Q} =\frac{T}{M}\frac{1}{r_{0}^{2}}\left[M^{2}-\
\S ^ { 2 }- 2M r_0 \right]\ . \end{equation}
In Fig.  6 the specific heat $C$ is plotted in
terms of ${M/M_{extr}}$ for the classical
Reissner-Nordstr\"om family of black holes ($P=Q$).  Note that the
specific
heat is positive near the extreme value of $M$ for all black holes
with $PQ
\not = 0$. Thus, as distinct from ordinary Schwarzschild black holes,
the
charged dilaton black holes  with $PQ \not = 0$ (as well as the
Reissner-Nordstr\"om ones) can be in a state of stable thermal
equilibrium
with `hot' matter with a temperature smaller than $T_{max} \sim
{1\over 8\pi
M_{extr}}$.

 The values of ${\cal P} = P/\sqrt 2 M$ and ${\cal Q} = Q/\sqrt 2 M$
for which $C^{-1}$ vanishes obey the equation
\begin{equation}
[{\cal  P} -{\cal Q}]^{4}-6[{\cal P}-{\cal Q}]^{2}
+8[{\cal P}+ {\cal Q}]-3=0\ .
\end{equation}
In the $(\cal{P},\cal{Q})$ plane, the line of the zeroes of
(\ref{C}) divides the square into two regions of different sign of
$C$.
This line is plotted in Fig.  7. There is also a line of zeroes of
$C$ in
the $(J,Q)$ plane in the case of Kerr-Newman black holes.  The fact
that
our black holes have both electric and magnetic charges is
analogous to the fact that Kerr-Newman black holes have angular
momentum.

It is clearly very important to understand the intrinsic properties
of
extreme black holes as candidates for the final state after the
process of
evaporation.  The extremal limit
$(r_+ - r_-) \rightarrow 0$ is also a limit when the supersymmetry is
restored since $r_0 = (r_+ - r_-)/2$ will be shown to be the
parameter of
supersymmetry breaking.   The understanding of supersymmetric
properties of extreme
black holes may shed  new light on the final stages of the
evaporation of
charged black holes.

 \section {Supersymmetric properties of dilaton black holes}
          \label{s:susy}
 First we will prove that non-extreme black holes always break
supersymmetry.
After that we will consider extreme solutions and find unbroken
supersymmetries.

In a bosonic background, the unbroken supersymmetries are determined
by
the terms depending on the bosons in the transformation laws of the
fermions.
Deleting there again the axion field, the relevant transformation
laws of the
$N=4$ theory are \footnote{We use the $SO(4)$ version for
convenience, but
the
duality transformation \eqn{dual} can be used to translate everything
to the $SU(4)$ version.} \cite{N4SO4,CSF,N4SC} (in chiral notation,
see Appendix~A)
\begin{eqnarray}
\half \d \Psi_{\m I} &=& \nabla _\m \e_I - {\textstyle\frac{1}{8}}
\s^{\r\s}T^+_{\r\s ,IJ} \g_\m  \e^J \ ,
\nonumber\\
\half\d \L_I &=&- \gamma^\mu \e_I \partial_\mu\phi
+ \frac{1}{\sqrt 2} \s^{\r\s}\left( e^{-\phi}  F_{\r\s}\alpha_{IJ} -
e^\phi \tilde G_{\r\s}\beta_{IJ}\right)^-\e^J \ , \end{eqnarray}
where the covariant derivative contains the spin-connection (see
Appendix~A), and would also contain a $U(1)$ connection if the axion
 had been included.
 $T_{\mu\nu , IJ}$ is an auxiliary field. Its
algebraic field equation, which was used to obtain the action
\eqn{actSU4N4}, has put it equal to \begin{equation}
T^+_{\mu\nu,IJ}=2\sqrt{2}e^{-\phi}\left(F_{\r\s} \alpha_{IJ}
+e^{2\phi}
\tilde
G_{\r\s}\beta_{IJ}\right)^+\ .\label{valT} \end{equation}

The local supersymmetry algebra of $N=4$ \cite{N4SC} contains the
following
terms
which are relevant for the solutions which we consider:
\begin{equation}
[\d_Q(\e),\d_Q(\e')]=-2\d_{g.c.}\left(
\bar\e'^I\g^a\e_{I}+h.c.\right)+\d_{Lor.}(\bar\e^I\e'^J
T^{+\,ab}_{IJ}
+h.c.)+\ldots\ ,\label{commutN4}
\end{equation}
where $\d_{g.c.}$ and $\d_{Lor.}$ are the general coordinate and
Lorentz
transformations which act on the vierbeins in the following way:
\begin{equation}
\d_{g.c.}(\xi) e_\mu^a= \partial_\mu\xi^a-\omega_\mu^{ab}\xi_b\
;\qquad \d_{Lor.}(\Lambda) e_\mu^a=\Lambda^{ab}e_{\mu b}\ .
\end{equation}
The non-vanishing value of the auxiliary field $T$ in our solution
will
imply that the last term of \eqn{commutN4} produces central charges.

To simplify the analysis of supersymmetry, we will use a
system of coordinates which has a conformally flat 3-dimensional
space. Defining
\begin{equation}
r = \r +
\frac{r_0^2}{4\r} + M\
\end{equation}
and $\rho ^2 =(x^i)^2$, we have
\begin{eqnarray}
ds^2 &=& e^{2U} dt^2 - e^{-2U+2C}(dx^i)^2 \ ,
\label{genmetC}\end{eqnarray}
where
\begin{equation}
e^C= \frac{\partial r}{\partial \rho }=1 - \frac{r_0^2}{4\r^2}\ .
\end{equation}

{}From \eqn{valT} we see that the only  non-zero components  of
$T_{\mu\nu,IJ}$  for purely electric $F$ and
$\tilde G$ (and no axion) are
 \begin{equation}
T_{\hat i\hat 0,IJ}=2\sqrt{2}\left(\alpha_{IJ}e^{-\phi}\partial_i
\psi
+\beta_{IJ} e^{\phi}\partial_i\chi\right)\ .
\end{equation}
We  now take the dilaton field going to a constant $\phi_0$ at
infinity, and
the electric field strengths going to zero as
\begin{equation}
\frac{x^i}{\rho }F_{\hat i\hat 0}=\partial_\rho
\psi=-\frac{Q_{elec}}{\rho ^2}+{\cal O}(\rho ^{-3}) \
,\qquad
\frac{x^i}{\rho }\tilde G_{\hat i\hat
0}=\partial_\rho \chi=-\frac{P_{elec}}{\rho ^2}+{\cal O}(\rho ^{-3})
\
 {}.
\end{equation}
Then, defining as before,
\begin{equation}
Q=e^{-\phi_0}Q_{elec} \ ,\qquad P=e^{\phi_0} P_{elec}\ ,
\end{equation}
we have at large $\rho $ the behavior
\begin{equation} T_{\hat \rho \hat 0,IJ}\equiv \frac{x^i}{\rho }
T_{\hat i\hat 0,IJ}=-2\sqrt{2}\,
\rho ^{-2}\left(\alpha_{IJ}Q+\beta_{IJ}
P\right) +{\cal O}(\rho ^{-3})\ .
\end{equation}
We will now take $T$ to have this asymptotic value for large
$\rho $.

We choose the antisymmetric matrices $\alpha_{IJ}$ and  $\beta_{IJ}$
 as
$\alpha^3$ and $\beta^3$ in the notation of \cite{GSO}.
\begin{equation}
\alpha_{IJ}=\pmatrix{0&1&0&0 \cr -1&0&0&0\cr 0&0&0&1\cr
0&0&-1&0\cr} \ ;\qquad
\beta_{IJ}=\pmatrix{0&-1&0&0 \cr 1&0&0&0\cr 0&0&0&1\cr
0&0&-1&0\cr}\ .\label{explalbe}
 \end{equation}
Thus, they are block diagonal in the pairs (1,2) and (3,4).
In each
of the pairs $T_{IJ}$ is proportional to $\epsilon_{IJ}$ , the
two-index
antisymmetric symbol with $\epsilon_{12}=1$  (or $\epsilon_{34}=1$).
So we
put
\begin{equation}
 T_{\hat \rho \hat 0,IJ}=-4\,\frac{Z}{\rho ^2}\, \e_{IJ}+{\cal
O}(\rho ^{-3})\ ,
\end{equation}
and the values of $Z$ in the $(1,2)$ and the $(3,4)$ sector are
respectively
\begin{equation}
z_1=\frac{Q-P}{\sqrt{2}} \ ,\qquad z_2=\frac{Q+P}{\sqrt{2}} \ .
\end{equation}
We see now that $Z$ is a central charge operation in the sense that
\begin{eqnarray}
[\d_Q(\e),\d_Q(\e')]&=&-2\d_{g.c.}\left(
\bar\e'^I\g^a\e_{I}+h.c.\right)\nonumber\\ &&
-4\frac{Z}{\rho ^2}\left( (\bar\e^I\e'^J
\e_{IJ})\d^+_{Lor.}+(\bar\e_{I}\e'_{J}
\e^{IJ})\d^-_{Lor.}\right)  +\ldots\ ,\label{commutZ}
 \end{eqnarray}
where $\d^+_{Lor.}$ is a selfdual version of a Lorentz
transformation,
 and $\d^-_{Lor.}$ an
anti-self-dual one, \begin{equation}
\d^\pm_{Lor.} =
\delta_{Lor.}\left(\Lambda^{ab}=\frac{x^i}{2r}\left(e^{\hat
i[a}e^{b]\hat 0}\pm\half \e^{abcd}e_c^{\hat i} e_d^{\hat
0}\right)\right)\ .
\end{equation}

Let us prove that {\it non-extreme charged dilaton black holes break
all supersymmetries}.

In the supersymmetry transformation law we will use now the metric
\eqn{genmetC}, and
the assumption that $F_{\mu\nu}$ and $\tilde G_{\mu\nu}$ are electric
solutions, i.e. that $F_{i0}=\partial _i \psi $ and $\tilde
G_{i0}=\partial
_i\chi $ are the only non-zero components. The covariant derivatives
on
spinors are \begin{equation}
\nabla_0 = \partial_0 - e^{2U-C}\partial_i U \sigma_{i0} \ ,\qquad
\nabla_i =
\partial_i -\sigma_{ij}\,\partial_j(U-C)\ , \end{equation}
where indices on derivatives are curved, while those on gamma
matrices are flat. The resulting transformation laws of the fermions
are
\begin{eqnarray}
\half \delta \Lambda_I &=&\gamma_i e^{-C}
\Bigl[e^{U}\, \partial_i\phi\cdot\epsilon_I
- \frac{1}{\sqrt 2} \left(\alpha_{IJ}\, e^{-\phi} \, \partial_i\psi -
\beta_{IJ}\, e^\phi\, \partial_i\chi\right)\gamma_0\, \epsilon^J
\Bigr] \ ,\nonumber\\
\half \delta \Psi_{0I}&=&\half\sigma_{0i}\,
e^{U-C}\Bigl[e^U\, \partial_i U\cdot\epsilon_I - \frac{1}{\sqrt 2}
\left(\alpha_{IJ}\, e^{-\phi}
\partial_i\psi +
\beta_{IJ}\, e^\phi\, \partial_i\chi\right)\gamma_0\epsilon^J
\Bigr] \ ,\nonumber\\
\half \d \Psi_{i I}&=&\partial_i\e_I-\sigma_{ij}\, \partial_j
(U-C)\cdot
\epsilon_I \nonumber\\ &&
-\frac{1}{2\sqrt{2}}
\gamma_j\gamma_i\Bigl[ \alpha_{IJ}\, e^{-\phi}\,
\partial_j\psi +
\beta_{IJ}\, e^\phi\, \partial_j\chi\Bigr]\gamma_0\, e^{-U}\epsilon^J
\ .
\end{eqnarray}

For preserved supersymmetries, the first two relations lead to
\begin{eqnarray}
\e_I\, \partial_i\, e^{U+\phi}&=&\sqrt{2}\,\g_0\,\alpha_{IJ}\, \e^J\,
\partial_i\p \ ,
  \nonumber\\
\e_I\, \partial_i\, e^{U-\phi}&=&\sqrt{2}\,\g_0\,\beta_{IJ}\, \e^J\,
\partial_i\chi  \ ,
\label{unbrsusy} \end{eqnarray}
while the remaining equation $\d \Psi_{iI}=0$ reduces to
\begin{equation}
\partial_i\e_I -\half\e_I\, \partial_i  U +\sigma_{ij}\, \epsilon
_I\, \partial
_j C =0\ . \end{equation}
Acting with  $\partial_j$ on this equation and antisymmetrizing
with respect to  $ij$ (the integrability condition) gives
zero only if $C$ is a constant, which proves that we can only have
supersymmetry for $r_0=0$, the `extreme case'.

Let us find now the {\it unbroken supersymmetries of extreme
charged dilaton black holes } with  $C= r_0 =0$. In that case we find
\begin{equation}
\epsilon_I= e^{\half U}\e_I^{(0)}\ ,\label{eps0}
\end{equation}
where $\e_I^{(0)}$ are constant spinors.

The relations \eqn{unbrsusy} (and consistency with their
complex conjugates in view of
$\alpha_{\phantom{}_{IJ}}\alpha^{JK}=-\delta_I^K$) imply that
\begin{equation}
\sqrt{2}(\psi-\psi_0)=a\,  e^{U+\phi} \ ,\qquad
\sqrt{2}(\chi-\chi_0)=b \,   e^{U-\phi}\ , \end{equation}
where $\psi_0$ and $\chi_0$ are undetermined constants, and
$a=-sgn(Q) $
and $b=-sgn(P)$.  We get now two conditions from eqs.
(\ref{unbrsusy}):
\begin{eqnarray}
\left(\epsilon_I-a\,
\alpha_{\phantom{}_{IJ}}\gamma_0\epsilon^J\right)
\partial_i(U+\phi)&=&0 \ , \nonumber\\
\left(\epsilon_I-b \,
\beta_{\phantom{}_{IJ}}\gamma_0\epsilon^J\right)
\partial_i(U-\phi)&=&0 \ .\label{releps412}
 \end{eqnarray}

Consider first the case in which both $PQ \neq 0$,
which means that neither $\partial_i(U+\phi)$ nor $\partial_i(U-
\phi)
$ vanishes.

We get in each of the four quadrants of Fig. 1 one unbroken
supersymmetry.
To see this, let us introduce a new basis for the supersymmetries,
\begin{eqnarray}\label{redef}
\e_\pm^{12} &=& \e_2 \pm \gamma_0 \e^1\ ,\qquad \e^\pm_{12} =\e^2 \pm
\gamma_0 \e_1\ , \nonumber\\
 \e_\pm^{34} &=& \e_4 \pm \gamma_0 \e^3\ ,\qquad \e^\pm_{34} =\e^4
\pm
\gamma_0 \e_3\ .
\end{eqnarray}
Note that these spinors are still chiral.

The unbroken supersymmetries in each quadrant are:
\begin{eqnarray}
\e_+^{34} \mbox{ and }   \e^+_{34}\ \ \mbox{ for } \ Q>0, P>0\
,\nonumber\\
\e_+^{12} \mbox{ and }  \e^+_{12}\ \ \mbox{ for } \ Q>0, P<0\ ,
\nonumber\\
\e_-^{34} \mbox{ and }  \e^-_{34}\ \ \mbox{ for } \ Q<0, P<0\ ,
\nonumber\\
\e_-^{12} \mbox{ and }  \e^-_{12}\ \ \mbox{ for } \ Q<0, P>0\ .
\end{eqnarray}
Thus, for each side of the square in Fig. 1 we have one unbroken
$N=1$ supersymmetry, each time a different part of the original
$N=4$ supersymmetry.

If $Q$ or $P$ are zero, then in the first case
$\partial_i(U+\phi)=0$, and in the second case
$\partial_i(U-\phi)=0$. So only
one
of the two conditions in (\ref{releps412}) for spinors applies, and
we
have
two remaining supersymmetries,
 \begin{eqnarray}
\e_+^{34}, \,  \e^+_{34}, \, \e_-^{12}, \, \e^-_{12}\ \ \mbox{ for }
\
Q=0, \,  P>0 \ ,   \nonumber\\
\e_-^{34}, \,   \e^-_{34},\, \e_+^{12}, \,
\e^+_{12} \ \ \mbox{ for } \ Q=0, \, P<0 \ , \nonumber \\
\e_+^{34}, \,   \e^+_{34},\, \e_+^{12}, \,
\e^+_{12}\ \ \mbox{ for } \ P=0, \,  Q>0  \ ,  \nonumber \\
\e_-^{34}, \,   \e^-_{34},\, \e_-^{12}, \,
\e^-_{12}\  \ \mbox{ for } \ P=0, \,  Q<0 \ .
 \end{eqnarray}
Thus in each vertex in Fig. 1 there is an unbroken $N=2$
supersymmetry,
each time a different part of the original $N=4$ supersymmetry.
And since the vertex of the square is the intersection of two sides,
the supersymmetries which are unbroken in the vertex are those which
are
unbroken on both sides adjoining the given vertex.

The bound $M\geq |Z|$ was derived in \cite{FSZ} using this basis for
the
spinors. Indeed, one may check that (with a sum over the pairs (1,2)
and
(3,4))
\begin{eqnarray} \bar\e'^I\g_0\e_{I}&=&\half\left(\bar\e'^+\g_0\e_+
+\bar\e'^-\g_0\e_-\right)\nonumber\\
 \bar\e'^I\g_i\e_{I}&=&\half\left(\bar\e'^+\g_i\e_-
+\bar\e'^-\g_i\e_+\right)\nonumber\\
\bar\epsilon'^I\bar\e^J\e_{IJ}&=&\for\left(\bar\e'^+\g_0\e_+ -
\bar\e'^-\g_0\e_-
+\bar\e'^-\g_0\e_+ - \bar\e'^+\g_0\e_-\right)\nonumber\\ &&
-(\e\leftrightarrow\e')\nonumber\\
\bar\epsilon'_I\bar\e_J\e^{IJ}&=&\for\left(\bar\e'^+\g_0\e_+ -
\bar\e'^-\g_0\e_-
-\bar\e'^-\g_0\e_+ + \bar\e'^+\g_0\e_-\right)\nonumber\\ &&
-(\e\leftrightarrow\e')\label{sgcm}\ .
\end{eqnarray}
The complex conjugates of the first two equations are $-$ these
expressions
with
$\e$ and $\e'$ interchanged.

This implies that when one takes the commutator as in \eqn{commutZ}
between two supersymmetries $\epsilon _+$ and $\epsilon '_+$ then
only
the space translation does not enter. The time translation is
proportional to the mass, and there is a central charge (acting as
$\d^+_{Lor.} +\d^-_{Lor.}$) proportional to Z. In this basis both
terms are proportional to $\bar \epsilon '^+\gamma _0\epsilon _+$.
The hermiticity properties then imply that $M+ Z$ should be
non-negative, and becomes zero only if $\epsilon _+$ is an unbroken
supersymmetry \cite{FSZ}. When one uses $\epsilon_-$, one can see in
\eqn{sgcm} that the sign
of the $Z$ contribution changes, and one finds $M-Z\geq 0$ and zero
only for unbroken $\epsilon _-$ supersymmetry. In this way one
obtains
$M\geq |Z|$. Using the pair (1,2) leads to $M\geq z_1$, while the
pair (3,4)
leads to the bound $M\geq z_2$. From these arguments it is also clear
that supersymmetries are unbroken if and only if the charges
are equal to their extreme values as in Fig.1.

There is nothing in our analysis which depends on spherical symmetry;
thus, the multi black hole case is included automatically.  We must,
however,
remember that $P$ and $Q$ refer to the total charge.   Only purely
magnetic or purely electric multi black holes possess $N=2$
supersymmetry;
a configuration with mixing between $P$ and $Q$, such as the one with
charges $(P,0)$ and $(0,Q)$, possesses only $N=1$ supersymmetry.

Thus we have shown that the extreme dilaton multi black hole
solutions
of $N=4$ supergravity, given in eqs. (\ref{H1H2}), (\ref{extsol}),
(\ref{nholes}), have some $N=1$ or $N=2$ supersymmetries unbroken.

Note that there exists a particular conformal transformation of the
original
canonical geometry  which brings the parameters of unbroken
supersymmetry
to  constant spinors, according to eq. (\ref{eps0}) :
\begin{equation}\label{conf}
\e_I^{(0)} = e^{-\half U} \e_I ^{can}  \ ,\quad
g_{\m\n} ^{(0)} = e^{-2 U} g_{\m\n}^{can} \ ,\quad
  \L ^{(0)}_I = e^{+\half U} \L_I  ^{can} , \quad \ etc \ .
\end{equation}
After such a conformal transformation, the metric takes the form
\begin{equation}
\label{time}
ds^{2 \,(0)} = dt^2 - e^{-4U} (dx^i)^2\ .
\end{equation}
In such a geometry, supersymmetry with parameter
$\e^{(0)}$ exists globally on  the space, in contrast with  the
canonical
geometry
where the parameter of unbroken supersymmetry is $ \e_I
^{can}=e^{\half U}
\e_I^{(0)}$ and goes to zero near the horizon. In addition, the time
component of the spin connection vanishes and the time derivative
coincides
with
the covariant time derivative. The conformal transformation
(\ref{conf})
 defines a choice of time coordinate for the supersymmetric state
for which the commutator of two supersymmetries is a translation
in time. For one of the solutions discussed above,
namely, for purely magnetic multi black holes with $U=-\phi$, the
corresponding conformal transformation is
\begin{equation}
g_{\m\n} ^{(0)} = e^{-2 U} g_{\m\n} ^{can} = e^{2 \phi} g_{\m\n}
^{can} = g_{\m\n} ^{string}\ .
 \end{equation}
 So we see that in the purely magnetic case the unbroken
supersymmetries
in the stringy geometry are realized in terms of {\it constant}
spinors.

\section{Nonrenormalization Theorem for the Partition Function of the
Extreme Charged Dilaton Black Hole}

The advantage of establishing the supersymmetric properties of
extreme
black holes is that of obtaining the possibility to prove a powerful
nonrenormalization  theorem for quantum corrections.  For classical
Reissner-Nordstr\"om extreme black holes,  the corresponding theorem
has
been established in \cite{K}; the analogous theorem exists for
extreme dilaton
black holes, as we will now show.

The language we will use for formulation of the theorem is that used
by
Gibbons and Hawking \cite{GHaw} for calculation of actions and
partition
functions in the context of black holes and de Sitter space.
In our previous publication \cite{K} we used  the language of the
``effective
on
shell action", which is often used  when considering  supersymmetric
theories.
It will be clear from the following equations that we will discuss
the
calculation of the same path integral as before, but for the black
holes, in addition, the thermodynamic interpretation  of the path
integral as
the partition function will be available.

The fundamental path integral in quantum gravity is:
\begin{equation}\label{PI}
Z = \int d[g] d[\Phi] \exp\{ i I[g, \Phi]\} \ ,
\end{equation}
where $d[g]$ is the measure on the space of metrics, $d[\Phi]$ is
the measure on the
space of matter fields and $I[g, \Phi]$ is the action. We assume that
the
path integral is
well-defined, i.e. an appropriate background invariant
gauge-fixing of all local symmetries
is performed. Let $g_0, \Phi_0$ be extremals of the classical action,
i.e. solutions of
classical equations of motion. One can then represent our integration
 variables as
\begin{eqnarray}
g&=& g_0 + \tilde g\ ,\\
\Phi &=& \Phi_0 + \tilde \Phi\ ,
\end{eqnarray}
and expand the action around this background
\begin{equation}
I[g,\Phi] = I[g_0,\Phi_0] + I_2[\tilde g, \tilde \Phi] + I_3[\tilde
g, \tilde
\Phi] +
\dots ,
\end{equation}
where $I_2$ contains terms quadratic in fluctuations, $I_3$ contains
terms
cubic in fluctuations, etc.

In other words, we are calculating the background functional by
expanding
it near the classical extremal (saddle point),
\begin{equation}\label{Z}
\ln Z = i I[g_0,\Phi_0] + \ln \int d[\tilde g] d[\tilde \Phi] \exp\{
i
(I_2[\tilde g,
\tilde \Phi] +
 + \dots) \} .
\end{equation}

At finite temperature, this path integral in Euclidean space can also
be
interpreted as a
thermal partition function with the properties
\begin{equation}
\ln Z = - I_{eucl} = -T^{-1} F\ ,
\end{equation}
where $F = M - TS  -  \sum_i \m_i C_i$ is the free energy
and $\m_i$ are chemical potentials associated with conserved charges
$C_i$, $S$ being the entropy of the system.
Gibbons and Hawking have calculated the classical
action on  the black hole solution in
\cite{GH} (the first term in Eq. (\ref{Z})) and in this way have
 established, for all black holes known at that time,  that
\begin{equation}\label{A}
-\ln Z^{cl} = I_{eucl} = S =  \for A \ ,
\end{equation}
i.e. the Euclidean action is $\for$ of the area of the horizon.
Our
investigation of electrically and magnetically charged dilaton black
holes
(\ref{sol2}) also confirms the rule (\ref{A}); the logarithm of the
partition function in the classical approximation is given by the
following
expression:

\begin{equation}\label{part}
-\ln Z^{cl} =  S =  \pi (r_+^2 - \S^2) = \pi ( [M + \half (r_+ -
r_-)] ^2 -
\S^2 ) \ .
\end{equation}

Our calculation of the value of the on-shell action required a
careful
treatment of all terms in the action, including the extrinsic
curvature term,
as was done in the calculation of the entropy of the
Reissner-Nordstr\"om
black hole in \cite{GHaw}.

The calculation of the partition function of one extreme spherically
symmetric dilaton black hole can be performed, for example, by
taking the limit $ r_+ \rightarrow  r_- $  in Eq. (\ref{part}). The
result is

\begin{equation} \label{vanish}
I_{extr}^{(1)} = -\ln Z_{extr}^{cl}= S_{extr} = \half \pi |z_2^2 -
z_1^2 |
= \pi ( M^2 - \S^2 )= 2 \pi |PQ|\ .
\end{equation}

Rather than taking the limit, we can calculate the action for extreme
purely
magnetic or purely electric black holes directly, even though the
temperature is not well defined. After we express the volume integral
of the Lagrangian as a surface integral, this term is exactly
cancelled by the surface integral of the extrinsic curvature. This
vanishing
of the Lagrangian, due to supersymmetry, confirms the previous result
(\ref{vanish}) at $PQ = 0$.

In the extreme case, the force balance condition is satisfied, so we
may have
multi black hole solutions which are not spherically symmetric.
However, we
can still calculate the partition function of such a  configuration,
by
following \cite{Brill}.  Using the equation of motion for the vector
fields,
we can write the action, including extrinsic curvature term, as a
single
surface integral.   Near the horizon of the $r$-th black hole, the
metric,
dilaton  and electromagnetic fields are all dominated by   terms
involving
only the charges of the  $r$-th black hole, so that the action ends
up being
just the
sum of $ \frac{1}{4}$ of the areas of the individual holes:
\begin{equation}\label{ttt}
I_{extr}^{(n)} = -ln Z_{extr}^{cl} = 2 \pi \sum_{r=1}^{n}  |P_r Q_r|
   =  \pi \sum_{r=1}^{n}   ({M_r}^2 - {\Sigma_r}^2 ) \ .
\end{equation}
Notice that the action $I_{extr}^n$ of the multi black hole
configuration is
always less
than that of the action $I_{extr}^{(1)}$ of a single black hole with
total
charges
$P =  \sum_{r=1}^{n}
|P_r| \neq 0$ and $Q=\sum_{r=1}^{n} |Q_r| \neq 0$:
\begin{equation}\label{mmulti}
I_{extr}^{(1)} = 2 \pi |P Q| = 2 \pi \sum_{r=1}^{n} |P_r| \sum_
{s=1}^{n} |Q_s|
= \pi (M^2 -\Sigma^2)\ .
\end{equation}
 Thus the total area of the horizons of the dilaton multi black hole
configuration
 at a given $P$ and $Q$ is smaller than that of one extreme
spherically
 symmetric black hole. All these solutions with any number of
black holes, but different Euclidean actions, have unbroken $N=1$
supersymmetry.

The extreme dilaton multi black holes
 have a special solution
 with only electric or only magnetic charge of each hole. All these
 solutions have zero Euclidean action, independently of the number
 of holes. This vanishing of the action is the consequence of the
higher
 unbroken supersymmetry, $N=2$, as opposite to the ones with
 mixing of $P$ and $Q$, which have only $N=1$ supersymmetry
 unbroken.

A nonrenormalization theorem can be derived in complete analogy with
that
in Ref. \cite{K} for the classical extreme Reissner-Nordstr\"om case.

For the extreme dilaton multi black holes the theorem can be
formulated
as follows:
The exact partition function of the extreme dilaton multi black hole
is
\begin{equation}
Z_{extr}^n = \exp \bigl(-\sum_{r=1}^n\for  A_r \bigr) \ ,
\end{equation}
 where $A_r$  is the area of the horizon of the $r$-th individual
hole, i.e.
{\it
the partition function, calculated in the semiclassical
approximation,
acquires no quantum
corrections.}

The absence of quantum corrections to the path
integral
$Z=exp(-I_{eucl}[g_0, \Phi_0])$  \, (\ref{Z}) takes place under the
following conditions. One should perform the calculations of the path
integral within
$N=4$ supergravity or in superstring theory. This means that
perturbations near the extreme dilaton black hole have to include the
graviton,
four gravitino, four dilatino, six vectors, a dilaton  and an
axion. Also
an
$N=4$ vector multiplet may be included.
 This means that the matter fields  $\Phi$ in Eq.  (\ref{PI})
are all fields which are superpartners of the graviton in the $N=4$
supermultiplet,
and a matter $N=4$ supermultiplet, including a gluon and gluino.
 There exists
a choice of the representation of the fields in the $N=4$ multiplet
which
leads to the absence of
one-loop conformal-axial anomalies \cite{Duff}.
 Also it is known that the one-loop
anomalies are absent in extended supergravities for $N\geq 3$
in the case that the loop calculations are performed in terms of
$N=1$
superfields. Starting from  $N=3$ supergravity, the net number of
chiral $N=1$ matter and  ghost
multiplets is zero, and therefore there are no anomalies. In
particular,
there is no divergent one-loop correction to the supersymmetric form
of the Euler number \cite{Duff}.

The proof of the nonrenormalization theorem for the on-shell
effective
action (\ref{Z})
consists of the following steps.

i)  Establishing that in $N=4$ supergravity or superstring theory
$Z[g_0, \Phi_0]$ has to be
a locally supersymmetric functional of supergravity fields for an
arbitrary
on-shell background.

ii) Realizing the property i)  in a form of manifestly supersymmetric
on-shell
$N=4$ superinvariants,
which are given by the integrals over the superspace.

iii)  Observing that manifestly supersymmetric on-shell $N=4$
superinvariants vanish in a bosonic background which has some
unbroken
supersymmetries. In those backgrounds the dependence on some
combinations
of Grassmann coordinates of the superspace vanishes and the
corresponding
superinvariants vanish due to the properties of Berezin integration
over the
anticommuting variables. For example, all local gauge-independent
counterterms  in this background acquire the form
\begin{equation}
\int d^4x d^{4N}\t \ \mbox {Ber}\, E  \ L(x, \t) = \int d^4x\
D_{unbr}
\Psi(x, \t)
|_{\t=0} \ ,
\label{inv} \end{equation}
where $\Psi$ is some spinorial superfield (the result of the action
of ($4N-1$)
fermionic derivatives
on the superfield Lagrangian).  The last fermionic derivative,
denoted by  $
D_{unbr}$, is chosen to correspond to one of the unbroken
supersymmetries of
the background. Using the fact that the  standard definition of the
supersymmetry variation of the superfield is
\begin{equation}
\d_{\e} \Psi (x, \t) =\sum_{\a =1}^{4N} \e^{\a} D_{\a} \Psi (x, \t) \
,
\end{equation}
we conclude that
\begin{equation}
 D_{unbr} \Psi(x, \t) |_{\t=0} =0\  ,
\end{equation}
and that the supersymmetric invariant (\ref{inv}) is vanishing in the
bosonic
background with an unbroken supersymmetry.

Thus, the existence of unbroken supersymmetries in the purely bosonic
background means that quantum corrections to the  partition function
cannot
change the semiclassical value of $\ln Z$ if the corrections satisfy
generalized
Ward identities following from local supersymmetry, i.e. if the
theory has no
anomalies.

The nonrenormalization theorem was derived above in the context of
the
Euclidean action.  However, our derivation of unbroken
supersymmetries in Sec. 6  was in the context of a space-time with
Lorentzian
signature.  One may wonder whether the theorem and the
supersymmetries
apply in both signatures.
The only previously known example of absence of quantum supergravity
corrections was in case of  super-self-dual instanton backgrounds
\cite{RK}.
These backgrounds exist only in Euclidean space; the unbroken
supersymmetry
and corresponding nonrenormalization theorem rely on properties of
Euclidean
space-time, where right-handed spinors can be set to zero while
left-handed
spinors may be non-vanishing . This does not extend to Lorentzian
space-time,
where right and left spinors are  complex conjugates of each other
and cannot
be
set to zero separately.
In contrast to this, the unbroken supersymmetries for extreme dilaton
black
holes exist in both
 signatures, since in deriving the constraints on the parameters of
unbroken supersymmetry we never used chirality properties specific to
Euclidean signatures.

Therefore our nonrenormalization theorem for extreme dilaton black
holes
holds for both  Euclidean and Lorentzian signatures.

\section{Discussion}

The main goal of this paper was to study the relationship between
black
hole  physics and supersymmetry. We were especially interested in
both
electrically and
magnetically charged dilaton black holes. Such black holes may appear
in
supergravity and in string theory.  They interpolate between purely
electric and  purely magnetic dilaton black holes, which exhibit very
interesting but somewhat confusing properties discussed by
many authors.

We found that supersymmetry indeed provides us with powerful tools
for
investigation of  black holes.
Firstly, we were able to find a supersymmetric theory which
contains electrically and magnetically charged
dilaton black holes. We have shown that the mass of each black hole
satisfies
 two
inequalities,  $M \geq |z_1|$ and  $M \geq |z_2|$, where $z_i$ are
the central
charges of the supersymmetry algebra, and are related to the electric
and
magnetic charges as follows: $z_1 =  \frac{Q -
P}{\sqrt 2}$ and $z_2 = \frac{Q + P}{\sqrt 2}$. When neither of these
inequalities is saturated (i.e. when $M > |z_1|$, $M > |z_2|$),
supersymmetry
is broken; when one is saturated, we have extreme
$N = 1$ supersymmetric dilaton black holes. The two inequalities can
be
saturated only for purely magnetic or purely electric extreme black
holes,
in which case $N=2$ supersymmetry becomes restored.

Thus, with the help of supersymmetry one can justify the very notion
of an
{\it extreme} black hole as a body which has a minimal mass for
given
values of charges. This implies that extreme black holes cannot
evaporate by emitting (uncharged) elementary particles. This is
consistent
with the vanishing of the Hawking temperature and/or breakdown of
the thermal description for extreme black holes. It is consistent
also with
the vanishing of the imaginary part of the effective action in the
extreme black hole background.

But the most unusual result, which we obtained with the aid of
supersymmetry, is  that the higher order quantum
supergravity or superstring corrections to the  effective  action
vanish in
both Lorentzian and Euclidean extreme dilaton black hole backgrounds.
Previously, the only example of such a background (with  Lorentzian
signature) was flat Minkowski space.  As a consequence of the
Euclidean
result, we were able to obtain an exact expression for the entropy of
the
extreme dilaton black hole.

This allowed us to describe properties of dilaton black holes near
the
extreme limit with greater confidence. We  calculated the
temperature,
entropy and specific heat of dilaton black holes as a function of
$M$, $P$
and $Q$. We have shown that even though the euclidean action
(entropy) for
extreme black hole can be calculated exactly, the usual
thermal description of dilaton black holes  breaks down near the
extreme
limit for all possible values of $P$ and $Q$.

Another interesting observation is the relation between supersymmetry
and
the cosmic censorship conjecture. Indeed, the supersymmetric bound on
the
black hole mass ensures that the black hole singularity is hidden by
the
event horizon. It approaches the event horizon only in the extreme
case when
$N = 2$  supersymmetry becomes restored, which is possible if only
one of the charges is present. It is not clear whether an evaporating
black
hole  can ever reach its extreme limit (due to breakdown of the
thermal
description), but even if it can do so, the singularity can never
appear {\it
outside} the event  horizon. This means that an outside observer will
never
see the singularity.

It is interesting that this relationship between supersymmetry and
the
cosmic  censorship hypothesis is valid for ordinary Schwarzschild
black
holes, for the Reissner-Nordstr\"om ones and, as we have shown in
this
paper, for a large class of electrically and magnetically charged
dilaton
black holes. This makes it very tempting to propose the following
Super Cosmic Censorship Conjecture :

1) Supersymmetry does not like naked singularities. It either hides a
singularity under the horizon, or keeps it at  the event horizon,
where it
still cannot be seen by an outside observer.

 2) Broken supersymmetry dislikes singularities even more.  When
supersymmetry is broken (which is the case in our universe), a
singularity
always  remains hidden under the horizon.

Throughout the paper, we have been studying the
supersymmetry of extreme black holes mainly in the canonical
geometry, i.e. with the metric of standard 4-dimensional Einstein
theory. It is, however, possible to address the question: what will
happen with unbroken supersymmetries after a conformal
transformation, for example to the stringy geometry? We have found a
very
nice feature of purely magnetic solutions: the string sees a geometry
which possesses
unbroken supersymmetry with {\it constant} spinors. It would be
interesting to understand why the magnetic dilaton black hole is
special in
this respect.

 Note that even though we embedded our black hole solutions in a
supersymmetric theory, all our solutions are purely bosonic.   It is
possible, therefore, that the positivity bounds which we obtained are
valid
for such bosonic solutions not only in the context of the
supersymmetric
theories where they were derived, but in other theories which have
the
same bosonic sector.
 A similar statement is known to be true in
$N=1$ supergravity, where it was shown that the positivity bound on
energy
in  supergravity implies the positivity of energy in ordinary gravity
\cite{DT,W}. As it was formulated by Grisaru \cite{DT}, it is enough
that Einstein theory
``knows" that it can be successfully coupled to gravitinos.

In this paper we discussed not only single extreme black holes, but
also
extreme multi
 black hole solutions. An important feature of these solutions is
that extreme black holes can be in an equilibrium state due to
cancellation
 of
Coulomb-like forces between their electric, magnetic, gravitational
{\it  and}
dilaton charges.
The multi black holes also have unbroken supersymmetry: $N=1$ for
$PQ\neq 0$ and $N=2$ for $PQ=0$, where $P, Q$ are the total charges
of the
multi black hole configuration.

 The existence of many equilibrium configurations of black holes
with
the  same total mass and charge, independent of the position of each
black
hole, raises many interesting questions. Is there any possibility of
quantum
tunneling between  these degenerate configurations, similar to the
tunneling
between vacua with different topological charges in QCD? Does this
degeneracy  mean that  the final result of a charged black hole
evaporation
will be not a single black hole but a quantum superposition of states
corresponding to  an arbitrary number of charged black holes with a
given
total mass and charge? Is it possible that at the last stage of the
evaporation
of
a charged black hole it splits into many smaller black holes? Our
understanding of these problems is rather limited. In order to
stimulate their investigation, we will discuss some relevant
issues in Appendix B.

The solutions which we presented depended on only two charges.  It is
natural to consider solutions with more parameters, for example those
discussed at the end of Sec. 3.   The study of the supersymmetry
properties
of those solutions is in progress. Also to be studied in this context
are dual
dilaton dyons and rotating black holes with or without axion.
Investigation of
these solutions would give us an extra opportunity to study the
relation
between
supersymmetry and the cosmic censorship conjecture.

In fact, our investigations suggest the following challenge. Is it
possible to
classify and to find {\it all solutions of Einstein theory,
interacting
with matter, which are supersymmetric?}
There exists a partial answer to this problem, given by Tod \cite{T},
for $N=2$ supergravity.  He has found {\it all metrics admitting
super-covariantly
constant spinors} of this theory. Currently, only part of his results
are
understood from the point of view of field theory, where the Einstein
and
Maxwell equations together with their right
hand side are derived from some Lagrangian.   Tod has solved
first order differential equations for unbroken supersymmetry, but
only
part of his solutions have been identified with solutions of some
covariant Lagrangian theory. For other supersymmetric theories
including
gravity,  such a complete analysis has not yet been performed ,
though
many interesting results are already known; see for example the
review
on supersymmetric string solitons in \cite{CHS} and the
results of the present paper for $N=4$ supergravity.

 Thus we expect on the basis of Tod's results that a rich family of
covariant Lagrangians and their solutions (not only asymptotically
flat spaces as studied here) might have a supersymmetric embedding in
the sense explained in our paper. Those theories will include plane
waves,
Israel-Wilson-Perjes metrics, etc. For all of these solutions we may
expect
that the unbroken supersymmetry will take quantum gravity corrections
under control.

{\bf Acknowledgments.}

The authors are grateful to D. Brill, M. Peskin, J. Russo, A.
Strominger,
L. Susskind and L. Thorlacius  for
numerous  fruitful discussions. We are very grateful to D. Linde  who
has
represented our results in a graphical form using the program
``Mathematica''.

The work of R.K., A.L.,  A.P. and A.V.P. was supported in part by NSF
grant
PHY-8612280. The work of  R.K.  was supported in part
 by the John and Claire Radway Fellowship in the School of
Humanities and Sciences at  Stanford University. The work of  A.P.
was supported in part by  Stanford University
Physics Department Fellowship Fund.
The work of T.O. was supported by a Spanish Government M.E.C.
postdoctoral
grant. A.V.P. thanks the Physics Department at Stanford University
for
the hospitality. His visit was also
supported by a travel grant of the N.F.W.O., Belgium.

\pagebreak

\section*{Appendix A. Notations and Conventions} \label{appnot}
We use the metric signature $(+---)$. The curved indices are  denoted
by
$\mu,\nu,\ldots  = 0,\ldots ,3$ and the flat ones  by $a,b,...$. If
restricted
to space,
we use $i,j,\ldots =1,2,3$ for both types of indices. From the
context it is
usually
clear which type they represent, e.g. on derivatives $\partial $ the
indices
are
curved, while on gamma matrices they are flat. Where confusion can
arise, we
use $\hat 0$ or $\hat i$ to indicate that indices are curved ones.

We define \begin{equation}
\epsilon^{\mu\nu\rho\sigma}=\sqrt{-g}\,e^\mu_a\,
e^\nu_b\, e^\rho_c \,e^\sigma_d \,\epsilon^{abcd} \ , \qquad
\epsilon^{0123}=i=-\epsilon_{0123}
\end{equation}
where the former implies that the latter is true for flat as as well
for curved indices. For spherical
coordinates we have $\epsilon ^{tr\theta \phi }=i$. The dual of an
antisymmetric  tensor is defined as
\begin{equation}
{} ^{\star} F^{ab}=\half \epsilon ^{abcd}F_{cd}\ .
\end{equation}
 We introduce the self-dual and anti-self dual tensors
\begin{equation}
F_{ab}^\pm =\half\left(F_{ab}\pm {}^\star F_{ab}\right) \ .
\end{equation}
If we write antisymmetric tensors as forms, there is the
correspondence
\begin{equation}
F =\half F_{\m\n} dx^{\m }\wedge dx^{\n}\ .
\end{equation}

Antisymmetrization is done with weight one: $[ab]=\half(ab-ba)$.
A symbol $\cdot$ is used to indicate that derivatives do not act
further to the right. Without such a symbol all $\partial
$-operations
are assumed to act on all fields to the right in the same terms,
unless it is enclosed in brackets.

The gamma and sigma matrices are defined by
\begin{equation}
\gamma_a \gamma_b=-\eta _{ab}+2\sigma_{ab} \ , \qquad
\gamma_5=i \gamma_0\gamma_1\gamma_2\gamma_3\ ,
\end{equation}
which implies that $\half\epsilon^{abcd}\sigma_{cd}=-\gamma^5
\sigma^{ab}$.
The matrices $\g_i$ and $\g_5$ are hermitian, while $\g_0$ is
antihermitian.

For the spinors, we use a chiral notation, where the chirality is
indicated by the position of the $I,J$ ($SO(4)$) index, or $\pm$
after
the redefinitions \eqn{redef}. Which position corresponds to which
chirality is not the same for all spinors. We have
\begin{equation}\begin{array}{ll}
\Psi_\mu^I=\half (1+\gamma _5) \Psi_\mu^I  \ , & \Psi_{\mu I}=\half
(1-\gamma _5) \Psi_{\mu I} \ , \\
\epsilon^I=\half (1+\gamma _5) \epsilon^I  \ , & \epsilon _I=\half
(1-\gamma _5) \epsilon_I \ , \\
\lambda^I=\half (1-\gamma _5) \lambda^I  \ , & \lambda _I=\half
(1+\gamma _5) \lambda_I\ . \end{array}\end{equation}
The conjugates for any spinor $\chi $ are
\begin{equation}
\bar \chi ^I=-i(\chi _I)^\dagger \gamma _0=(\chi ^I)^T{\cal C} \ ,
\end{equation}
where $\cal C$ is the charge conjugation matrix
\begin{equation}
{\cal C}^T=-{\cal C} \ , \qquad {\cal C}\g_a{\cal C}^{-1}=-\g_a^T\ ,
\end{equation}
such that e.g. $\bar \epsilon^I\gamma _5=\bar \epsilon ^I$. In chiral
notations, the antisymmetric tensors are often automatically
(anti)self-dual. For example,
\begin{equation}
\sigma^{ab}F_{ab}\epsilon^I=\sigma^{ab}F_{ab}^-\epsilon^I\ .
\end{equation}

For the spin connections and curvatures we have
\begin{eqnarray}
\omega_\mu{}^{ab}&=&2e^{\nu[a}\partial_{[\mu}e_{\nu]}{}^{b]}
-e^{a\rho}e^{b\sigma}
e_{\mu c}\partial_{[\rho}e_{\sigma]}^c \ , \nonumber\\
R_ {\mu\nu}^{ab}(\omega)&=&2\partial_{[\mu}\omega_{\nu]}{}^{ab}
+2\omega_\mu^{c[a}\omega_\nu{}^{b]}{}_c \ , \nonumber\\
R_{\mu\nu}&=& e_a{}^\rho e_{\mu b}R_{\nu\rho}{}^{ab} \ , \qquad
R=-R_{\mu\nu}g^{\mu\nu}\ . \label{Rconv}\end{eqnarray}
This implies that
\begin{equation}
\frac{\delta }{\delta g^{\mu\nu} }\int d^4x\sqrt{-g}R =
-\sqrt{-g}\left(R_{\mu\nu}+\half g_{\mu\nu}R\right)\ .
\end{equation}
The covariant derivative on spinors is
\begin{equation}
\nabla_\mu\epsilon=
\left(\partial_\mu-\half\omega_\mu{}^{ab}\sigma_{ab}\right)\epsilon\
{}.
\end{equation}

For the translation from \cite{N4SO4,CSF} (the notations are in the
first article) we
replaced their $\gamma_a$ by $i\gamma_a$ and $\gamma_5$ gets a minus
sign. We replace their $\epsilon^{\mu\nu\rho\sigma}$ by
$-i\epsilon^{\mu\nu\rho\sigma}$. For the spinors, their $\Psi _\mu $
becomes $\frac{1}{\sqrt{2}} \Psi_\mu$ and $\epsilon$ becomes
$\sqrt{2}\epsilon$, and we changed the sign of $A$.

For translation from \cite{N4SC} we changed the metric, thus
$g_{\mu\nu}$ gets a minus sign, at any
implicit appearance, e.g., in $\partial^\mu$.  The vierbein $e_\mu^a$
is unchanged, but then, e.g., $e_{\mu a}$ gets a minus sign.
Note that with the translation given above, $\omega$ is the same as
in \cite{N4SC} which is opposite to the conventions of \cite{PvNPR}.
But we define $R_{\mu\nu}$ and $R$
such that they are the same (the minus sign in the last equation in
\eqn{Rconv} is thus due to the metric).
\newpage

\section*{Appendix B. Splitting of Extreme Black Holes}

A very interesting situation appears when we consider multi black
hole
solutions. As we already noted, they describe an equilibrium
configuration
of black holes which have the same mass and total charge independent
of the
position of each individual black hole.  Let us now try to understand
the
behavior of such a configuration at the quantum level.   One may
expect that
quantum fluctuations of the metric, as well as those of the dilaton
and vector
fields, may lead to quantum jumps of the positions of black holes. In
a normal
situation, when the total energy of a system depends on the positions
of its
constituents, this would lead to some  change $\Delta E$ of the
energy of the
system which would violate energy conservation. Such a system then
returns
to its original state within a time $(\Delta E)^{-1}$, in accordance
with
the uncertainty principle. However, in our case $\Delta E = 0$ for
any change
of the extreme black hole configuration. This suggests that extreme
black
holes  behave as Brownian particles at a flat surface. If originally
they are
localized in some place, later on the distance between them grows and
eventually become indefinitely large. This behavior has a simple
quantum
mechanical interpretation as a  spreading of the wave packet
describing
several noninteracting  particles.

Now we can make a second step and ask the question: what will happen
if an
extreme black hole splits quantum mechanically into two extreme black
holes of the same total mass
and charge? This process is not forbidden by energy and charge
conservation.
Usually, it is forbidden by the second law of black hole physics,
since in
such a process the total area of the horizons of the black holes (and
the total
entropy) would decrease. However, the thermodynamic interpretation of
the
law  suggests that this process may be possible due to fluctuations
of the
entropy $\Delta S < 0$, even though the probability of such
fluctuations will
be exponentially suppressed. Moreover, as distinct from the ordinary
Reissner-Nordstr\"om black holes, the area of the horizon (and the
entropy)
of purely magnetic and purely electric extreme {\it dilaton} black
holes
vanishes. Thus, the second law of black hole physics does not forbid
their
splitting.

To obtain an intuitive (though, admittedly, vague) understanding of
the
process of splitting, let us consider a purely electric or magnetic
dilaton
black hole near its horizon. The horizon is singular (for the moment,
we will
not consider the stringy version of a magnetic black hole), and the
black hole
can be completely described by a conformastatic metric
(\ref{genmet}), with
the singularity at $x = 0$. Since the area of the horizon vanishes,
one may
imagine that quantum fluctuations of the metric on the Planck scale
can easily
split the singularity into two, i.e. we will get a conformastatic
metric with
two black holes very close to each other. (After all, we cannot
actually
interpret processes on the Planck scale in terms of classical space
time with a
fixed number of classical singularities.) In a normal situation, such
an event
would not have any interesting consequences, since the baby black
hole would
immediately recombine with its parent. However,
extreme black holes described by the conformastatic solution
(\ref{genmet}),
(\ref{extsol}) do not attract each other. If our picture of Brownian
motion of
extreme black holes is correct, then the average distance between the
baby
black holes and their parent can only grow. This is very similar to
the
standard picture of black hole evaporation: If the black hole is not
surrounded by ultrarelativistic particles with a large temperature,
the
particles emitted by the black hole move away and its mass decreases.
Similarly, if the universe is not filled by a dense gas of black
holes, then
the
new black holes, produced by the black hole splitting,
typically will move away due to  Brownian motion. Perhaps a more
adequate
way to say it is to remember that splitting of the black hole  (which
changes the  number of singularities) occurs without any energy
release.
Thus, the
products of splitting have  vanishing relative momenta, which means
that the
distances between them become  indefinitely large. In this sense, the
theory
of the black hole splitting  resembles the theory of baby universe
formation,
where the baby universe  is produced with vanishing energy and
momenta,
hence the place where it is created  cannot be localized.  Another
useful
analogy is the tunneling between different vacua in QCD which have
the same
energy but are characterized by different topological numbers.

In a more general case, when the black hole has both electric and
magnetic
charges, its singularity is hidden under the horizon. Then the
transition
between one black hole, with one singularity, and two black holes,
with two
singularities separated by their two horizons, is discontinuous. This
is why
we expect the probability of such processes to be exponentially
suppressed.
To get an estimate of the probability of splitting, one may use
standard
thermodynamic arguments, which suggest that it should be proportional
to
$\exp (\Delta S)$, where $\Delta S$ is the change of entropy
\cite{Brill}.\footnote{This argument was used in \cite{Brill} applied
to splitting of Bertotti-Robinson universes, which have the same
geometry as
the geometry near the horizon of the Reissner-Nordstr\"om black
hole. It is interesting that this simple argument sometimes gives
correct
results even in some situations where the description of tunneling in
terms of
instantons is ambiguous; for example, it gives a correct expression
for
the probability of tunneling in an inflationary universe, in terms of
the
entropy of de Sitter space.}  If this interpretation is correct, one
may expect
that the probability of splitting of one Reissner-Nordstr\"om black
hole into
many, with a total mass $M = \sum_r  {M_r}$, is given by
\begin{equation}\label{spl1}
 \Gamma \sim e^{- \pi \left( (\sum_r  {M_r})^2 -   \sum_r  {M_r^2
}\right)} \ .
\end{equation}
One can easily see that the probability of splitting of large black
holes is
exponentially suppressed. However, this suppression may be not very
strong
for small black holes with masses of the order of the Planck mass
$M_p = 1$.

An analogous expression for dilaton black
holes, which follows from Eqs. (\ref{ttt}), (\ref{mmulti}) in Sec. 7,
is
\begin{eqnarray}
 \Gamma \sim&& e^{- \pi \left( (\sum_r  {M_r})^2 - (\sum_r
{\Sigma_r})^2 -
\sum_r  ({M_r}^2   - {\Sigma_r}^2 )\right)}\nonumber\\ && = e^{ -
2\pi
\left( \sum_r |P_r|
\sum_s |Q_s| - \sum_r |P_r Q_r|  \right)}
= e^{-2 \pi \left( \sum_{r \not = s} |P_r| |Q_s| \right)}\ .
\label{spl2}
\end{eqnarray}
This expression,
unlike Eq. (\ref{spl1}), shows that there is no exponential
suppression
of splitting of purely electric or purely magnetic dilaton black
holes. This
agrees with our qualitative discussion of  quantum fluctuations and
Brownian
motion of extreme black holes. It would be very desirable to confirm
(or
disprove) the validity of Eqs. (\ref{spl1}), (\ref{spl2}) by finding
an
explicit instanton solution which is
responsible for the black hole splitting.  It  may well happen that
the
suppression of splitting is given by a more complicated expression
than $\exp
(\Delta S)$, especially in the situation where the thermal
description of black
holes breaks down. However, at this stage it would be most important
to
understand whether this probability is finite at all. If this is the
case (and
at least for purely electric black holes it seems to be a reasonable
possibility), then the physics of black holes may prove to be even
more
interesting than we thought.
\vfill
\pagebreak

\section*{Figure captions}

Figure 1.   The space of electrically and magnetically charged
dilaton
black holes with charges $P$ and $Q$ and a fixed mass $M$.
\, $Q_{max}=P_{max}=\sqrt{2}M$.
Every point inside the square corresponds to a regular black hole.
The points outside the square (which are forbidden by supersymmetry)
correspond to metrics with naked singularities. The points {\it on}
the square
correspond to extreme black holes. The unbroken $N = 1$
supersymmetries
for the extreme black holes on each of the four sides (I, II, III,
IV) of the
square are shown. In the corners, we have unbroken $N = 2$
supersymmetry.

Figure 2.  Isothermals in the space of charged dilaton black holes of
constant
mass. The interval of temperature between two contiguous isothermals
is $\frac{1}{50} T_{max}$, where $T_{max} = \frac{1}{8\pi
M}$ is a temperature of the Schwarzschild black hole with a mass $M$.
The two axes of coordinates are isothermals corresponding to
$T=\frac{1}{8\pi M}$. The four sides of the square (excluding the
corners)
 are isothermals corresponding to $T=0$. The corners are very
special:
 All the isothermals (for all the different allowed
temperatures) converge to the corners. This can be better seen in
Fig. 3.

Figure 3. The temperature of charged
 dilaton black holes of a given mass $M$  as a function of $z_1/M =
(Q -
P)/\sqrt 2 M$ and $z_2/M =  (Q + P)/\sqrt 2 M$. The extreme black
holes
correspond to the sides $|z_1|/M = 1$ and $|z_2|/M = 1$ of the
square.

Figure 4.  The temperature versus the mass for different electric and
magnetic charges. For definiteness, we take $Q > P > 0$. The black
hole
evaporates until its mass approaches the limiting value $M_{extr} =
z_2 = (Q +
P)/\sqrt 2$. Three families of curves correspond to $z_2 = 1$, $z_2 =
2$
and $z_2 = 4$.  For each of these values of  $z_2 = (Q + P)/\sqrt 2$
we
choose three different $P/Q$ ratios: $P/Q=1,\, \frac{1}{4},\, 0$. The
smoothest curves are the ones with $P=Q$ (classical
Reissner-Nordstr\"om). The sharpest correspond to the limit $P/Q
\rightarrow 0$, which reproduces purely electric dilaton black holes.
There is always a maximum for the temperature (a point where the
specific heat diverges and reverses sign), and always the temperature
falls sharply to zero in the vicinity of the bound. This implies the
breakdown of the thermal description when we approach extremality
{\it
for all values of} $P$ {\it and} $Q$.

Figure 5.  The entropy $S$ of the charged dilaton black holes as a
function of
$z_1/M $ and $z_2/M$. It has a
maximum for the Schwarzschild black hole, which corresponds to  the
origin
of coordinates ($P=Q=\Sigma=0$).  For purely electric
or magnetic extreme dilaton black holes (in the corners) it is zero.
On the sides of the square temperature vanishes, but the total
entropy
(euclidean action) remains non-zero, $S = 2\pi\,|PQ|$.

 Figure 6. The specific heat of a  Reissner-Nordstr\"om black
hole with $P = Q > 0$ as a function of $M/M_{extr} = \sqrt 2 M/(Q +
P)$.

Figure 7. The locus of points in the $({\cal P}, {\cal Q})$ plane
where the
specific heat of a fixed-mass charged dilaton black hole diverges.
Inside the
curve, the specific heat is negative; outside, it is positive.
\pagebreak


\begin{thebibliography}{100}
\bibitem{GH} G.W. Gibbons and C.M. Hull, Phys. Lett.  {\bf
109B},  190  (1982).
\bibitem{GG} G.W. Gibbons, in: {\it Supersymmetry, Supergravity and
Related Topics},  eds. F. del Aguila, J. de Azc\'arraga and L.
Ib\'a\~nez
(World Scientific,  Singapore 1985), p. 147.
\bibitem{K} R. Kallosh, Supersymmetric Black Holes, Stanford
University preprint SU-ITP-92-1, to be published in Phys. Lett. {\bf
B}.
 \bibitem{Duff} M. Duff, in: {\it Supergravity' 81}, eds. S. Ferrara
and J.G.
Taylor (Cambridge University Press 1982).  S. Gates, M.
Grisaru, M. Ro\v cek and W. Siegel,  {\it Superspace}
(Benjamin/Cummings,
London 1983), p. 451.
\bibitem{G} G.W. Gibbons, Nucl. Phys. {\bf B207},  337 (1982).
\bibitem{GM} G.W. Gibbons and K. Maeda  Nucl. Phys. {\bf B298},  741
(1988).
\bibitem{GHS}  D. Garfinkle, G.T. Horowitz  and A. Strominger,
 Phys. Rev {\bf D43}, 3140 (1991).
\bibitem{HW}
J. Preskill, P. Schwarz, A. Shapere, S. Trivedi and F. Wilczek,
Mod. Phys. Lett.  {\bf A6}, 2353 (1991).
C. F. E. Holzhey and F. Wilczek, Black holes as
elementary particles,  preprint IASSNS-HEP-91/71, December 1991.
\bibitem{STW} A. Shapere, S. Trivedi and F. Wilczek,
 Mod. Phys. Lett.  {\bf A6}, 2677 (1991).
\bibitem{B} W.B. Bonnor, Gen. Rel. Grav. {\bf 12}, 453 (1980).
\bibitem{PenHawk} R. Penrose, {\it Structure of Space-Time}, in:
Battelle
Rencontres, 1967 Lectures in Mathematics and Physics, edited by C. M.
DeWitt
and J. A. Wheeler (Benjamin, New York,1968);
S.W. Hawking and G.F. Ellis, {\it The Large Scale Structure of
Space-Time},
Cambridge University Press, Cambridge (1973).
\bibitem{Pen1} R. Penrose, Rev. Nuovo Cimento, {\bf 1}, 252 (1969).
\bibitem{Pen2} R. Penrose, in: {\it General Relativity, an Einstein
Centenary
Survey}, eds. S.W. Hawking and W. Israel (Cambridge Univ. Press,
Cambridge
1979).
\bibitem{Wald} R. Wald, {\it General Relativity}, (Univ. of Chicago
Press, Chicago 1984).
\bibitem{WO} E. Witten and D. Olive, Phys. Lett. {\bf 78B},  97
(1978).
\bibitem {FSZ} S. Ferrara, C.A. Savoy and B. Zumino, Phys. Lett.
{\bf 100B}, 393 (1981).
\bibitem{DT} S. Deser and C. Teitelboim,  Phys. Rev. Lett. {\bf 39},
249
(1977);
M.  Grisaru, Phys. Lett. {\bf 73B}, 207 (1978).
 \bibitem{W} E. Witten, Comm. Math. Phys. {\bf 80}, 381
(1981);  J.M. Nester, Phys. Lett. {\bf 83A}, 241  (1981);
 W. Israel and J.M. Nester, Phys. Lett.    {\bf 85A}, 259  (1981);
 C.M. Hull, Commun. Math. Phys. {\bf 90}, 545 (1983).
\bibitem{N4SO4} A. Das, Phys. Rev. {\bf D15}, 2805 (1977);
E. Cremmer, J. Scherk and S. Ferrara, Phys. Lett. {\bf 68B}, 234
(1977);
E. Cremmer and J. Scherk, Nucl. Phys. {\bf B127}, 259 (1977).
\bibitem{CSF} E. Cremmer, J. Scherk and S. Ferrara
 Phys. Lett.  {\bf 74B},  61  (1978).
\bibitem{dual} S. Ferrara, J. Scherk and B. Zumino, Nucl. Phys. {\bf
B121}, 393 (1977);
E. Cremmer, J. Scherk and S. Ferrara, Phys. Lett. {\bf 74B}, 61
(1978);
B. de Wit,  Nucl. Phys. {\bf B158}, 189 (1979);
E. Cremmer and B. Julia, Nucl. Phys. {\bf B159}, 141 (1979);
M.K. Gaillard and B. Zumino, Nucl. Phys. {\bf B193}, 221 (1981).
\bibitem{T} K.P. Tod, Phys. Lett. {\bf 121B}, 241 (1983).
\bibitem{HH} J.H. Horne and G.T. Horowitz, Rotating Dilaton Black
Holes,
Preprint UCSB-TH-92-11, 1992.
\bibitem{N4SC} E.  Bergshoeff, M.  de Roo and B.  de Wit, Nucl.
Phys. {\bf B182}, 173 (1981);
M.  de Roo, Phys.  Lett.  {\bf 156B}, 331 (1985); Nucl.  Phys.
{\bf B255}, 515 (1985).
\bibitem{GSO} F.  Gliozzi, J.  Scherk and D.  Olive, Nucl.  Phys.
{\bf B122}, 253 (1977).
\bibitem{GHaw}G.W. Gibbons, S.W. Hawking, Phys. Rev.
{\bf D15}, 2752 (1977).
\bibitem{Brill} D. Brill, Splitting of an Extremal Reissner-
Nordstr\"om
Throat via Quantum Tunnelling, Preprint UDM-92-176 (1992).
\bibitem{RK} R.E. Kallosh, JETP Lett.  {\bf 29} (1979) 172 and 449;
Nucl. Phys. {\bf B165},  119 (1980);
 S. Christensen, S. Deser, M. Duff and M. Grisaru, Phys. Lett.
{\bf 84B},  411 (1979).
\bibitem{CHS} C.G. Callan,  J.A. Harvey and A. Strominger,
Supersymmetric
String Solitons,  preprint EFI-91-66 (1991).
\bibitem{PvNPR} P. van Nieuwenhuizen, Phys. Rep. {\bf 68}, 189
(1981).
\end{thebibliography}
\end{document}